\documentclass[aps,amssymb,twoside,twocolumn,showpacs,showkeys]{revtex4}

\usepackage[pdftex]{graphicx}
\usepackage{amsfonts}
\usepackage{amsmath}
\usepackage{color}
\usepackage{rotating}
\usepackage{multirow}

\newcommand{\be}{\begin{equation}}
\newcommand{\ee}{\end{equation}}
\newcommand{\bea}{\begin{eqnarray}}
\newcommand{\eea}{\end{eqnarray}}

\newcommand{\brn}{\nonumber \\ \nobreak}
\newcommand{\ign}[1]{} 

\newcommand{\alphaC}{\alpha_\mathrm{C}}
\newcommand{\alphaP}{\alpha_\mathrm{P}}

\newcommand{\cola}[1]{\colorbox[rgb]{0.7,1.0,0.7}{\raisebox{0pt}[4.5pt][0pt]{#1}}}
\newcommand{\colb}[1]{\colorbox[rgb]{0.9,0.9,0.0}{\raisebox{0pt}[4.5pt][0pt]{#1}}}
\newcommand{\colc}[1]{\colorbox[rgb]{1.0,0.5,0.0}{\raisebox{0pt}[4.5pt][0pt]{#1}}}
\newcommand{\cold}[1]{\colorbox[rgb]{1.0,0.0,0.0}{\raisebox{0pt}[4.5pt][0pt]{#1}}}
\newcommand{\sidew}[1]{\multirow{2}{*}{\begin{sideways}#1\end{sideways}}}

\begin{document}

\title{On optimal wavelet bases for the realization of microcanonical cascade processes}

\author{Oriol Pont}
\email{opont@ub.edu}
\affiliation{Departament de F\'{\i}sica Fonamental. Universitat de Barcelona. Diagonal, 647. 08028 Barcelona. Spain}

\author{Jose M.D. Delgado}
\email{josemaria.delgado@uam.es}
\affiliation{Departamento de F\'{\i}sica Te\'orica. Universidad Aut\'onoma de Madrid. Campus de Cantoblanco. 28049 Madrid. Spain}

\author{Antonio Turiel}
\email{turiel@icm.csic.es}
\affiliation{Institut de Ci\`encies del Mar - CSIC. Passeig Mar\'{\i}tim de la Barceloneta, 37-49. 08003. Barcelona. Spain}

\author{Conrad J. P\'erez-Vicente}
\email{conrad@ffn.ub.es}
\affiliation{Departament de F\'{\i}sica Fonamental. Universitat de Barcelona. Diagonal, 647. 08028 Barcelona. Spain}

\date{July 29th, 2008}

\begin{abstract}
Multiplicative cascades are often used to represent the structure of multiscaling variables in many physical systems, specially turbulent flows. In processes of this kind, these variables can be understood as the result of a successive transfer in cascade from large to small scales. For a given signal, only its optimal wavelet basis can represent it in such a way that the cascade relation between scales becomes explicit, i.e., it is geometrically achieved at each point of the system. Finding such a basis is a data-demanding, highly-complex task. In this paper we propose a formalism that allows to find the optimal wavelet in an efficient, less data-demanding way. We confirm the appropriateness of this approach by analyzing the results on synthetic signals constructed with prescribed optimal bases. Although we show the validity of our approach constrained to given families of wavelets, it can be generalized for a continuous unconstrained search scheme.
\end{abstract}


\maketitle

\section{Introduction}

Multiplicative processes giving rise to cascades are quite ubiquitous in Nature. Either as a real mechanism or as an effective one, cascades spontaneously develop in many scale-free systems. For instance, in a three-dimensional flow under fully developed turbulence, the energy is transferred from large to small scales (where it is released) through a well defined cascade process. In analogy, for an arbitrary cascade, the transfer of information between different scales, measured in terms of appropriate variables, reveals also an interesting hierarchical structure \cite{Frisch.1995} whose analysis provides useful information about some features of the system. In particular the distribution of scaling exponents can be determined from the study of the statistical properties of the cascade process \cite{Chhabra.1989,Turiel.2006}.

Apart from some models \cite{Schertzer.1987,Benzi.1993a,Chainais.2007}, there have been few attempts to characterize the structure of particular signals in terms of local cascade descriptors. The advantage of such an approach is that one can extract geometrical information about the signal in contrast to standard methods where only global statistical information is available. Given a signal (or dataset), the key point is to find a representation basis where the cascade process can be expressed in a \emph{microcanonical} form, in other words, to find an appropriate transformation in which the representation variables are precisely these local cascade descriptors.

Wavelets are a standard analysis tool in signal processing \cite{Mallat.1999}: wavelet projections are integral transforms that separate the relevant details of a signal at different scale levels, and since they are tuned to an adjustable scale, they are appropriate to analyze the multiscale behavior of cascade processes and to represent them. Most of the standard wavelets are able to accurately estimate the distribution of energy (or equivalent quantity) at each stage of the cascade, something that is very useful as a global descriptor. In addition, for a given system, there is a particular wavelet called {\it optimal wavelet} that also characterizes the dynamics at a local level, as it corresponds to the proper representation basis for cascade variables. The main advantage of optimal wavelet projections is that they can be expressed as products of successive cascade variables chosen along a branch of a dyadic tree. This representation is minimally redundant, as cascade variables are independent between consecutive cascade stages, and it defines a local effective dynamics that opens the way to new theoretical developments and practical applications \cite{Buccigrossi.1999,Turiel.2003a,Pottier.2008}.

An attempt to find the optimal wavelet of natural images from a sample dataset has been reported in \cite{Turiel.2000c}. The methodologies presented there are quite limited, as they exploit particular symmetries of natural images, and the uncertainty in the so-obtained empirical optimal wavelet is rather large to allow fine developments. In this paper we will present an improvement of the methodology presented in \cite{Turiel.2000c} in order to derive the optimal wavelet of more general types of data with more precision. Our study is focused on theoretical and methodological aspects of this problem, and is validated using synthesized data with known optimal wavelet.

The paper has the following arrangement: The next section explains the concept of multiplicative cascade and how it is identified in real signals. In section~\ref{sec:cascades} we mathematically formalize canonical and microcanonical cascades through the use of wavelet projections, and we also introduce the concept of optimal wavelet. In section~\ref{sec:optimization} we introduce a quantifier of the optimality degree and discuss about optimization strategies. Then, we generate synthetic cascades and check their optimality, showing the results in section~\ref{sec:results}. Finally, in section~\ref{sec:conclusions} we give our conclusions.

\section{Persistence in scale invariant signals}
\label{sec:persistence}

Multiplicative cascades are present in many different systems, but they are not usually recognized as such. Usually, their presence is reported by means of indirect evidence about its effects on the properties of signals. One of the most commonly reported effect of multiplicative cascades is the persistence of feature detection across scales. The importance of persistence is that the detection of a feature at a coarse scale allows to infer the presence of the same feature at finer scales. This phenomenon is well known since the introduction of wavelet representation of signals, and it is first described by Mallat and co-workers \cite{Mallat.1989,Mallat.1991}. The optimal wavelet is the one that maximizes this inference capability.

To understand what is the role of wavelet processing it is convenient to clarify what a multiresolution decomposition is.
In a multiresolution decomposition the signal can be represented as a combination of wavelet coefficients that can be arranged according progressive levels of resolution, from finer to coarser. This representation is just an algebraic change of basis, so the multiresolution decomposition of a signal contain exactly the same information as the original signal, and we can pass from one to the other with a linear transformation and without any loss of information.  In the case of 1D signals a single wavelet can be used to fully represent the signal in a dyadic scheme; for 2D signals, we need three different wavelets that will expand three different pyramids of resolution levels. In a dyadic scheme, when we pass from one resolution to the next coarser one the scale changes by a factor two, i.e., the diameter of the wavelet at the coarser scale is exactly twice the diameter of the wavelet at the previous, finer scale. This implies that a wavelet coefficient obtained at the coarser scale affects an area that is twice larger in diameter than that of the finer scale; roughly speaking, a wavelet coefficient at the coarser scale covers the area of two wavelet coefficients at the finer scale in 1D and the area of four wavelet coefficients at the finer scale in 2D. In section~\ref{sec:cascades} the concepts of wavelet basis and dyadic decomposition will be introduced in greater detail; see also \cite{Daubechies.1992,Mallat.1999}.

\begin{figure}[htb]
\begin{center}
\leavevmode
\includegraphics[width=4cm]{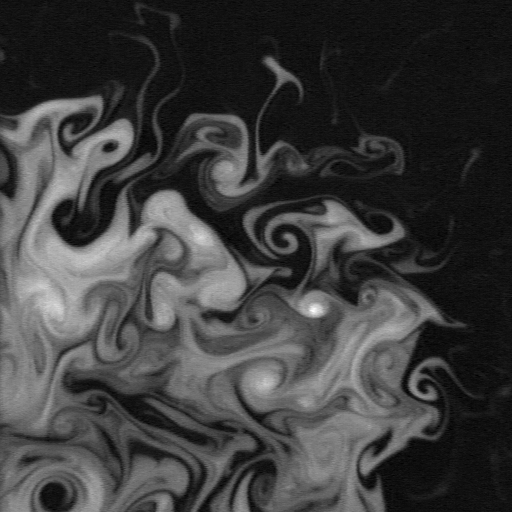}
\includegraphics[width=4cm]{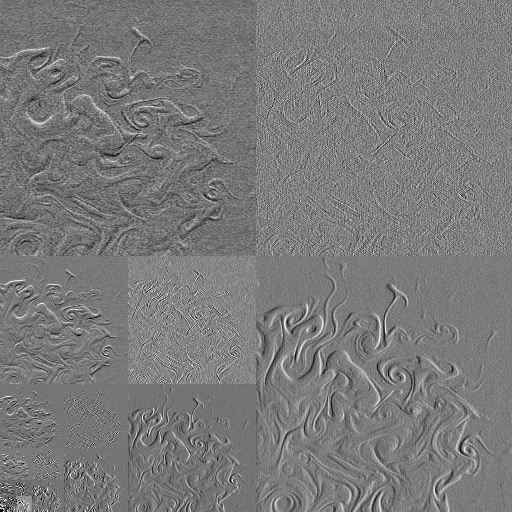}
\end{center}
\caption{{\bf Left:} Snapshot of dye distribution submitted to the action of 2D turbulence; see the text for details. {\bf Right:} Multiresolution decomposition with Haar basis of the image on the left; each resolution level and orientation has been independently normalized to enhance details.}
\label{fig:multires}
\end{figure}

In Figure~\ref{fig:multires} we show a typical example of edge persistence. In the left panel we present a CCD-recorded snapshot of the distribution of dye under the action of 2D turbulence; the image was obtained in a laboratory experiment of dispersion of passive scalars under the action of direct enstrophy cascade (for details on the experiment see \cite{Paret.1998,Paret.1999,Jullien.2000}). In the right panel we show a multiresolution decomposition of this image in a 2D separable wavelet basis, namely Haar basis. The multiresolution decomposition on the right panel of Figure~\ref{fig:multires} presents all the wavelet coefficients of the representation in a compact shape. A 2D multiresolution basis requires three wavelets and hence there are three types of wavelet coefficients, which in this case can be labeled as horizontal (leftmost squares), vertical (those with a side on the bottom of the panel) and diagonal.

\begin{figure}[htb]
\begin{center}
\leavevmode
\includegraphics[width=2.6cm]{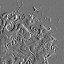}
\includegraphics[width=2.6cm]{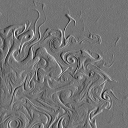}
\includegraphics[width=2.6cm]{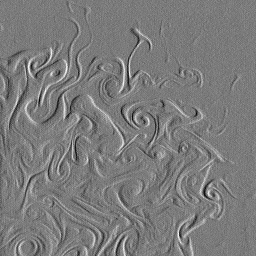}
\end{center}
\caption{The three finer resolution levels of vertical wavelet coefficients, extracted from Figure~\ref{fig:multires}; going from left to right we go from the coarser to the finest resolution. The three resolution levels are represented at the same size to help comparison.}
\label{fig:multires_detail}
\end{figure}

In Figure~\ref{fig:multires_detail} we present a detail of three consecutive resolutions of vertical coefficients extracted from Figure~\ref{fig:multires}. Notice that the multiresolution decomposition is just a change of vectorial basis, so the wavelet coefficients are algebraically independent. It is however obvious from Figure~\ref{fig:multires_detail} that the coefficients do not take arbitrary values: the edges detected at coarser scales persist at the same location but with better resolution at the finer scales. This is the persistence of edges, and it is a consequence of the structure of the signal, which implies that on many real systems edges are multiscale.

Edge persistence is a strong, relevant feature of physical signals, as it implies that the signal is highly redundant. It is precisely by means of the wavelet representation that this redundancy becomes evident. Persistence implies that we can predict to some extent what is going to happen at the next resolution level from the wavelet coefficients of a given level. Some authors \cite{Buccigrossi.1999,Wainwrigth.2001} have exploited this redundancy to devise algorithms for image compression. Particularly, Simoncelli and co-workers have noticed that the mutual dependence between consecutive scales can be better highlighted using conditional histograms \cite{Huang.1999,Buccigrossi.1999,Wainwrigth.2001,Schwartz.2001}. The histograms of fine-scale (also called ``child'') coefficients conditioned by the value of the coarse-scale (also called ``parent'') coefficient at the same location have a clear tie-bow shape for any wavelet \cite{Buccigrossi.1999,Schwartz.2001} (we also observe the same behavior in Figure~\ref{fig:illushisto} top). This shape implies that the dispersion of the child increases with the absolute value of the parent coefficient. This suggests that the child coefficient depends on its parent coefficient in a multiplicative fashion. For that reason, the distribution of the logarithm of the child coefficient conditioned by a value of the logarithm of the parent coefficient exhibits a linear dependence \cite{Buccigrossi.1999,Huang.1999} (see also our Figure~\ref{fig:illushisto} bottom). The authors found that, depending on the wavelet, the range of validity of this linear dependence can be larger or smaller.

More recently, Pottier et al. \cite{Pottier.2008} studied satellite images of surface chlorophyll concentration and found them to be persistent across scales. Although they used very different wavelet bases, for none of them the histogram of the logarithm of the child conditioned by the logarithm of the parent have a full linear range. As we will see later, a wavelet for which the conditioned histogram is fully linear is called optimal, in the same sense that the one introduced by Turiel and Parga in \cite{Turiel.2000c}. Pottier et al. proposed a particular model to describe the child-parent dependency, valid for many different wavelets that are not the optimal one but are not too far from it anyway. We will call this model the linear model, and it reads as:
\be
\alphaC = \eta_0 \, \alphaP + \alpha_0
\label{eq:suboptimal}
\ee

\noindent
where $\alphaC$ stands for the child wavelet coefficient and $\alphaP$ stands for its parent (i.e., it is obtained at the immediately coarser scaler and at the same position). $\eta_0$ and $\alpha_0$ are random variables mutually independent, also independent from $\alphaP$. The authors observed that this model fits reasonably well the conditioned histograms for many different wavelet bases, although depending on the particular basis the amplitude of the variable $\alpha_0$ varies; for smaller $\alpha_0$ the linear range in the conditioned histogram is larger and the converse. Now a question reasonably raises: is there any particular choice of wavelet for which the amplitude of $\alpha_0$ vanishes? This would be the optimal wavelet, in the same sense as in \cite{Turiel.2000c,Turiel.2003a}.

The importance of finding such an optimal wavelet must be stressed. First, because with the aid of this wavelet the description of the mutual dependence between parents and children can be simplified; in fact, $\alpha_0=0$ implies that the mutual information between $\alphaP$ and $\alphaC$ is maximized. So, a coding scheme as the one proposed by \cite{Buccigrossi.1999} attains the highest quality and smallest coding cost with the use of this wavelet. Besides, using this wavelet basis the inference of the value of the coefficients is improved, what has an impact on the quality of reconstruction algorithms to fill data gaps (as in \cite{Pottier.2008}) or on forecasting of time series. Finally, optimal wavelets can be used to derive improved models of multifractal systems (for instance, some variables under fully developed turbulence).

\section{Towards an optimal representation of data}
\label{sec:cascades}

\subsection{Canonical cascades}

The paradigm of systems in which multiplicative cascades develop are scale-invariant systems with one or many fractal interfaces. In them, conveniently designed intensive variables put in evidence a complex interplay between different scales.

Let $s(\vec{x})$ be a physical variable representing the signal under study. To study the scale relations of the system, we will need a properly defined, intensive, scale-dependent functional $\mathbf{T}$ applied to the signal, $\mathbf{T}[s](r, \vec{x})$. This variable depends on the point $\vec{x}$ and a scope or scale parameter $r$ that characterizes the range of influence of the functional. Typical examples of such a functional include the derivative at radius $r$, nonlinear measures based on the derivative or wavelet projections.

The canonical approach to multiplicative cascades is a statistical approach. Hence, the object under study is the distribution of the variable $\mathbf{T}[s](r, \bullet)$ for different values of the scale parameter $r$ only, disregarding the localization $\vec{x}$, i.e., considering all the points as statistically equivalent. That is why we will simply denote this variable as $\mathbf{T}_r$. The analysis of its distribution is achieved through its order-$p$ moments; studying the moments is enough to completely define the distribution provided they do not diverge too fast with $p$ \cite{Carleman.1922}.

A multiscaling (also called multifractal) signal $s$ is characterized by the power-law scaling in the order-$p$ moments of the related variable $\mathbf{T}_r$, in the way:
\be
\langle \mathbf{T}_r^p \rangle \; = \; A_p^{\mathbf{T}} \, r^{\tau_p} \: + \: o (r^{\tau_p})
\label{eq:anomalous}
\ee

\noindent
Recall that the symbol $o (r^{\tau_p})$ means a contribution that is negligible compared to $r^{\tau_p}$ when $r$ goes to zero.
In fractal signals, the exponent $\tau_p$ is directly proportional to the moment order $p$ and the proportionality constant is called singularity exponent or Hurst exponent. In multifractal signals \cite{Frisch.1995}, the dependence of $\tau_p$ on $p$ is nonlinear, a fact known as \emph{anomalous scaling}. In \ref{app:connection}, the connection between geometry and statistics of multifractal signals is discussed in greater detail.

In order to separate the part of the statistics that has to do with changes in scale, two different scales $r,L$ with $r<L$ can be compared, so:
\be
\langle \mathbf{T}_r^p \rangle = \left( \frac{r}{L} \right)^{\tau_p} \langle \mathbf{T}_L^p \rangle
\label{eq:relmomp}
\ee

\noindent
which is valid at lowest order in the limit of small $r$ and $L$. For some particular $\tau_p$, this relation implies the existence of a variable $\eta_{\kappa}$ such that:
\be
\langle \eta_{\kappa}^p \rangle = {\kappa}^{\tau_p}
\label{eq:etamoments}
\ee

\noindent
where $\kappa = r/L < 1$. Notice that one of the conditions for the existence of this variable is the validity of the expansion above, which in turn depends on taking a scale ratio parameter $\kappa$ smaller than 1; for this reason we have taken the ratio of the smaller scale by the larger scale. Notice also that there is no general proof on the existence of $\eta_{\kappa}$ for an arbitrary $\tau_p$; it can however be assumed to exist if $\tau_p$ defines infinitely divisible processes \cite{Novikov.1994,She.1994,She.1995,She.2007}. These cases cover many situations of interest, such as log-normal, log-L\'evy or log-Poisson processes.

With the aid of the variable $\eta_{\kappa}$ we can express eq.~(\ref{eq:relmomp}) in a more elegant way, making the cascade relation explicit:
\be
\mathbf{T}_r \doteq \eta_{r/L} \, \mathbf{T}_L
\label{eq:doteta}
\ee

\noindent
with $\eta_{r/L}$ and $\mathbf{T}_L$ being mutually independent. Here the symbol $\doteq$ means that the equality holds distributionally, i.e., $\rho(\mathbf{T}_r) = \rho(\eta_{r/L} \, \mathbf{T}_L)$. However, this relation does not necessarily hold pointwise, as we will explain in the following subsection.

The introduction of eq.~(\ref{eq:doteta}) now allows to split the statistics of the scaling variable $\mathbf{T}_r$ in two parts: one part, given by $\eta_{r/L}$, accounts for the properties of transformation under changes in scale, while the other part, given by $\mathbf{T}_L$, takes into account the behavior at a given reference scale $L$. Taking $L$ as the largest possible scale in the system, the distribution of all the variables $\mathbf{T}_r$ at any arbitrary scale $r$ can be referred to the fixed level $\mathbf{T}_L$ once the process of change in scale, $\eta_{r/L}$, is known.

We will call the $\eta_{r/L}$ \emph{cascade variables}. Their distributions do not depend on the particular scales $r$ and $L$ they connect but only on the scale ratio $\kappa=r/L$. If we now consider three scales $r<r'<L$ and we apply eq.~(\ref{eq:doteta}) to the three possible scale pairs it follows:
\be
\eta_{r/L} \doteq \eta_{r/r'} \, \eta_{r'/L}
\label{eq:cascade}
\ee

\noindent
from which the name ``cascade variable'' becomes evident: the variable relating scales $r$ and $L$ is equivalent to the product of the variables relating any two intermediate scales. If any intermediate scale is allowed, it follows that the cascade variables must have an infinitely divisible distribution \cite{Novikov.1994,Dubrulle.1994,Castaing.1996}. Another important characteristic of the distribution of the cascade variables is that it is a property of the signal and does not depend on the particular functional $\mathbf{T}$ used to obtain them, i.e., any functional capable to resolve the scaling exponents $\tau_p$ of the signal in eq.~(\ref{eq:anomalous}) leads to exactly the same distribution of cascade variables $\eta_{r/L}$ \cite{Frisch.1995}.

\subsection{Microcanonical cascade}
\label{subsec:microcascade}

Equation (\ref{eq:doteta}) makes sense only as a distributional equality and does not imply that the functional of scale $r$ at some point $\vec{x}$ is related to the functional of scale $L$ at the same point through an independent multiplicative factor. In general, $\mathbf{T}[s](r, \vec{x})$ and $\mathbf{T}[s](L, \vec{x})$ are not related by a variable $\eta_{r/L}(\vec{x})$ that is statistically independent of $\mathbf{T}[s](L, \vec{x})$. Of course, we can always define $\tilde{\eta}_{r/L}(\vec{x})$ as the ratio of these two variables,
\be
\tilde{\eta}_{r/L}(\vec{x}) \; = \; \frac{\mathbf{T}[s](r, \vec{x})}{\mathbf{T}[s](L, \vec{x})}
\label{eq:microepsiloncascade}
\ee

\noindent
but for most of the possible functionals $\mathbf{T}$, the variables $\tilde{\eta}_{r/L}(\vec{x})$ are not independent of $\mathbf{T}[s](L, \vec{x})$ and thus they cannot be considered cascade variables, as they do not verify eq.~(\ref{eq:cascade}). It is convenient to deal with cascade variables, as they are independent of the starting scale and only depend on the ratio of scales; this implies that they serve both to characterize the global properties of the system and to compactly codify its dynamics.

In many multifractal systems, the cascade process governs their dynamics as a local effective mechanism, what implies that there is a local variable $\eta_{r/L}(\vec{x})$ transferring energy, matter or information (depending on the system) from coarser to finer scales. Therefore, there may exist a system variable $s$ and a scale-tunable functional $\mathbf{T}$ for which eq.~(\ref{eq:doteta}) makes sense not only distributionally but also at any point $\vec{x}$ of the system. That is what we call \emph{microcanonical cascade}.

Among the functionals that are most commonly used to analyze the scaling properties of multifractal systems, wavelets occupy a prominent position. Wavelets have been used to perform local Fourier analysis and to characterize the local singularities of functions \cite{Mallat.1992}. In many different multifractal systems, wavelet projections have been used to characterize their scaling properties with success \cite{Turiel.2000a,Turiel.2003b}. Something that is very convenient about wavelet projections is that they can be inverted to retrieve the original signal \cite{Daubechies.1992}, so wavelet projections do not only analyze the signal, but also constitute a representation of it. That is why wavelet projections are good candidates to realize the microcanonical cascade.

A {\it wavelet} is a function that oscillates in the center of its domain and decays in its tails; we can think about wavelets as a pulse that decays very fast. Let $s(\vec{x})$ be a multifractal signal and let $\Psi(\vec{x})$ be a wavelet. We define the wavelet projection of $s$ on $\Psi$ at the position $\vec{x}$ and the resolution scale $r$ as:
\be
\mathbf{T}_{\Psi}[s](r, \vec{x}) \; \equiv \; \int \! \mathrm{d} \vec{y} \: s(\vec{y}) \: \Psi \!\! \left(\frac{\vec{x}-\vec{y}}{r}\right)
\label{eq:def_wavprojection}
\ee

\noindent
In terms of wavelet projections, a microcanonical cascade has the following form:
\be
\mathbf{T}_\Psi [s] (r, \vec{x}) \; = \; \eta_{r/L}(\vec{x}) \ \mathbf{T}_\Psi [s] (L, \vec{x}) 
\label{eq:optcascade}
\ee

\noindent
Notice that the key point is that $\eta_{r/L}(\vec{x})$ has to be both a cascade variable --in the sense of eq.~(\ref{eq:cascade})-- and independent from $\mathbf{T}_\Psi [s] (L, \vec{x})$. We can thus define the optimality of a wavelet as the degree of independence of $\tilde{\eta}_{r/L}(\vec{x})$ vs. $\mathbf{T}_\Psi [s] (L,\vec{x})$; we will discuss this possibility in depth in Section~\ref{sec:optimization}. There are evidences that such an optimal wavelet exists in natural images \cite{Turiel.2000c} and in marine turbulence \cite{Pottier.2008} for the specific case of wavelet dyadic representations.

\subsection{Dyadic representations of the cascade}
\label{subsec:dyadic}

Wavelet projections of a signal can be used to characterize the local properties of the signal \cite{Mallat.1992} or to represent it in an efficient scheme \cite{Daubechies.1992}. Although a signal can be retrieved from its wavelet projections if the wavelet is {\it admissible} \cite{Daubechies.1992}, such a representation is highly redundant and so a subset of wavelet projections must be retained. A typical way to subsample continuous wavelet projections is to select a dyadic subset (then called {\it wavelet coefficients}) from which the signal is fully reconstructed. In a dyadic subset, scales vary by a factor two and at each resolution level the positions are taken as integer amounts of the resolution size.

To keep formulae simple, hereafter we will limit our attention to one-dimensional systems. Hence, given a 1D signal $s(x)$ and a wavelet $\Psi$ capable to spawn a dyadic representation basis, the signal can be expanded as a series of wavelet terms:
\be
s(x) \; = \; \sum_{j=-\infty}^{\infty} \sum_{k} \alpha_{j, k} \, \Psi_{j, k}(x)
\label{eq:wv_expansion}
\ee

\noindent
where
\be
\Psi_{j, k}(x) = 2^{-j/2} \: \Psi(2^{-j}x-k)
\ee

\noindent
and $j, k$ are integer numbers. The coefficients of this representation basis, the $\alpha_{j, k}$, are called \emph{wavelet coefficients}. As $\Psi$ defines a representation basis, there is a unique set of wavelet coefficients $\{ \alpha_{j, k} \}$ such that eq.~(\ref{eq:wv_expansion}) is valid. The $2^{-j/2}$ normalization factor ensures that the 2-norm is 1, $\int \! \mathrm{d} x \, |\Psi_{j, k}(x)|^2=1$.

If the wavelet basis is orthonormal, i.e.,
\be
\left\langle \Psi_{j, k} \| \Psi_{j', k'} \right\rangle \; \equiv \; \int \! \mathrm{d} x \, \Psi^\ast_{j, k}(x) \, \Psi_{j', k'}(x) \; = \; \delta_{j, j'} \, \delta_{k, k'}
\label{eq:orthonormal}
\ee

\noindent
we can obtain the wavelet coefficients as projections on the wavelet basis, namely:
\be
\alpha_{j, k} \; = \;  \left\langle \Psi_{j, k} \| s \right\rangle
\label{eq:ortho_coefficients}
\ee

\noindent
If the wavelet basis is not orthonormal, the extension is rather straightforward: each basis vector $\| \Psi_{j, k} \rangle$ has its dual $\langle \tilde{\Psi}_{j, k} \|$ so that $\langle \tilde{\Psi}_{j, k} \| \Psi_{j', k'} \rangle = \delta_{j, j'} \, \delta_{k, k'}$ and the wavelet coefficients can be obtained as $\alpha_{j, k} = \langle \tilde{\Psi}_{j, k} \| s \rangle$.

In terms of a dyadic representation, the cascade takes a relatively simple form. For any wavelet basis, the canonical cascade relation, eq.~(\ref{eq:doteta}), takes the following form:
\be
\alpha_{j, k} \; \doteq \; \eta_{\frac{1}{2}} \: \alpha_{j-1, \left\lfloor {k}/{2} \right\rfloor}
\label{eq:dyadic_cascade}
\ee

\noindent
where the notation $\left\lfloor {k}/{2} \right\rfloor$ means \emph{the integer part of} ${k}/{2}$. Here we have written the cascade relation mimicking eqs.~(\ref{eq:doteta}) and (\ref{eq:optcascade}), although that the wavelet coefficients $\alpha_{j, k}$ are not intensive variables as the wavelet projections are as defined in eq.~(\ref{eq:def_wavprojection}) (while wavelet projections are $\infty$-norm normalized, wavelet coefficients are 2-norm normalized, which is highly convenient in the derivations to follow, especially in section~\ref{sec:optimization}). This means that the $\eta$-like variables written hereafter will differ from those appearing in Equations~(\ref{eq:etamoments})~to~(\ref{eq:optcascade}) in a constant normalization factor of $\sqrt{{r}/{L}} = \frac{1}{\sqrt{2}}$.

Notice that $\alpha_{j, k}$ is the wavelet projection at the scale $r_\mathrm{C}=2^{j}$ and position $x_\mathrm{C}=2^{j} \, k$, while $\alpha_{j+1, \left\lfloor {k}/{2} \right\rfloor}$ is the wavelet projection at the coarser scale $r_\mathrm{P}=2^{j+1}$ and position $x_\mathrm{P}=2^{j+1} \left\lfloor {k}/{2} \right\rfloor$; the positions $x_\mathrm{C}$ and $x_\mathrm{P}$ differ at most by $r_\mathrm{C}$, which is the spatial uncertainty at the scale $r_\mathrm{P}$, so at the scale $r_\mathrm{P}$ we can consider that $x_\mathrm{C}$ and $x_\mathrm{P}$ refer to the same position. To alleviate the notation, for given fixed scale index $j$ and position index $k$, $\alphaP \equiv \alpha_{j+1, \left\lfloor {k}/{2} \right\rfloor}$ is known as the Parent coefficient, $\alphaC \equiv \alpha_{j, k}$ is the Child coefficient and the cascade variable is $\eta \equiv \eta_{\frac{1}{2}}$, and we just write the canonical cascade relation above as:
\be
\alphaC \; \doteq \; \eta \: \alphaP
\ee

A dyadic wavelet basis is said to be optimal if the associated wavelet coefficients verify the microcanonical cascade relation, namely:
\be
\alphaC \; = \; \eta \: \alphaP
\label{eq:optimal}
\ee

\noindent
where $\eta$ is independent of the parent wavelet coefficient $\alphaP$ and is thus a cascade variable with associated scale ratio $\frac{1}{2}$.

It has been shown that if the optimal wavelet exists, there is a constructive formula to unambiguously obtain it from a large enough dataset \cite{Turiel.2000c}. This formula proves uniqueness of the wavelet, but it is rather unstable, specially in estimating the tails of the wavelet, and it is not very useful unless a large amount of data is available as learning set \cite{Turiel.2004a,Delgado.2006}. We will next analyze alternative strategies for a more stable determination of the optimal wavelet.

\section{Optimization from suboptimal representations}
\label{sec:optimization}

\subsection{Quadrature Mirror Filters}

Dyadic wavelet expansions can be used to describe the cascade with a discrete set of parameters. A particularly, widely used way to implement dyadic representation is in terms of Quadrature Mirror Filters (QMF), which are very robust in practical applications.
The main advantage of QMFs is that they are discrete filters so both obtaining the wavelet coefficients from a signal and reconstructing the signal from its wavelet coefficients are numerically exact operations (apart from round-off errors).
In the following, we will summarize the most relevant facts about QMFs; the interested reader can consult some wavelet textbooks \cite{Daubechies.1992,Mallat.1999}.

When a function $\Psi$ defines a wavelet basis, it is possible to find another function $\Phi$, called unity function, that is orthogonal to the wavelet but, contrarily to it, has non-zero mean. Another important property of unity functions is that they can be used to represent the approximation of the signal at a given scale.

The approximation of a signal $s(x)$ at a scale indexed as $j_0$ is given by an expansion of functions $\Phi_{j_0, k}$ whose coefficients are called \emph{approximation coefficients}. The signal can hence be expanded as a series of infinite levels $j$, as in eq.~(\ref{eq:wv_expansion}), or partially expanded up to a level $j_0$ and approximated at this level. Namely, we can expand the signal $s(x)$ as follows:
\bea
s(x) &=& \sum_{j=-\infty}^{\infty} \sum_k \alpha_{j, k} \, \Psi_{j, k}(x)
\\
     &=& \sum_{j=-\infty}^{j_0} \sum_k \alpha_{j, k} \, \Psi_{j, k}(x) \: + \: \underbrace{\sum_k \beta_{j_0, k} \, \Phi_{j_0, k}(x)}_{A_{j_0}(x)}
\nonumber
\eea

\noindent
The approximation $A_{j_0}(x)$ can be expressed either as a wavelet expansion of all the levels coarser than $j_0$ or as an expansion on unity functions at a single level $j_0$. Hence, it is possible to obtain the wavelet coefficients from the approximation, as the wavelet projections of the approximation coincide with those of the signal at any level coarser than $j_0$. It should be noticed that if the signal is discrete, it coincides with its approximation at any level finer than that of the discretization scale.

The main advantage of this new decomposition is that the approximations $A_{j_0}(x)$ are numerable sums, so we can define two numerable filters, denoted by $\{ g_n \}$ and $\{h_n \}$, that can be used to obtain the wavelet coefficients at any level provided that we know the approximation at the finest level, i.e., the signal at its discretization level. Then, applying the conjugate mirror filters $\{ g_n \}$ and $\{ h_n \}$ we can both obtain the wavelet coefficients from the signal or retrieve the signal from their coefficients with very fast algorithms, which are exact over discretized collections of coefficients.

When we expand the scaling function itself $\Phi$ (i.e., $\Phi_{0, 0}$) up to the next coarser scale $j_0=1$, the filter $\{ g_n\}$, which we will denote by the vector $\vec{g}=(\ldots,g_{-1},g_0,g_1,g_2,\ldots)$, is given by the wavelet coefficients, while the filter $\{ h_n \}$, which we will denote by the vector {$\vec{h}=(\ldots, h_{-1},h_0,h_1,h_2,\ldots)$}, is given by the approximation coefficients, namely:
\be
\Phi(x) \; = \; \sum_n g_n \, \Psi_{1, n}(x) + \sum_n h_n \, \Phi_{1, n}(x)
\ee

\noindent
where the wavelet coefficients at any level finer than $j=1$ are zero because the unity function coincides with itself at level $j=0$.

Let us now suppose that we have a discretized signal $s_k$ defined by a collection of values, which are naturally identified as the approximation coefficients at the highest resolution $\beta_{0, k} = s_k$. We will denote this collection of approximation coefficients by the vector $\vec{\beta}_0=(\ldots,\beta_{0, -1},\beta_{0, 0},\beta_{0, 1},\beta_{0, 2},\ldots)$. Since we have previously said that $r=2^j$, having the highest resolution at level $j=0$ means that we are expressing $r$ in units of pixels. To obtain the wavelet coefficients at the next coarser level $j=1$ we apply the filter $\vec{g}$. Let $\vec{\alpha}_{1}$ be the vector of these wavelet coefficients, then we have:
\be
\alpha_{1, k} \; = \; \sum_n g_{n - 2 k} \, \beta_{0, n}
\label{eq:wv_expansion1}
\ee

\noindent
that is, the filter $\vec{g}$ acts by convolution on the vector ${\vec{\beta}}$. For later convenience, let us introduce the matrix $\mathbb{G}$ that represents the action of $\vec{g}$ by convolution, i.e., $\mathbb{G}_{nn'}=g_{n' - 2 n}$. We can now elegantly express eq.~(\ref{eq:wv_expansion1}) in vectorial form as:
\be
{\vec{\alpha}}_{1} \; = \; \mathbb{G} \cdot {\vec{\beta}}_0
\label{eq:finest_wav_coeff}
\ee

Notice that the expression above can be used to relate the approximation and the wavelet coefficients of any two consecutive resolution levels, i.e.,

\be
{\vec{\alpha}}_{j+1} \; = \; \mathbb{G} \cdot {\vec{\beta}}_{j}
\ee

\noindent
but in order to obtain the wavelet coefficients at any other resolution we need an expression to obtain the coarser approximations derived from the highest resolved one. This can be done by means of the filter $\vec{h}$. Analogously to what has been derived previously, we have that two consecutive approximation levels can be related by the filter $\vec{h}$ as follows:
\be
{\vec{\beta}}_{j+1} \; = \; \mathbb{H} \cdot {\vec{\beta}}_{j}
\ee

\noindent
where $\mathbb{H}_{nn'}=h_{n' - 2 n}$. We already have the essentials to perform a perturbative analysis on the wavelet.

\subsection{Perturbative analysis}

In general, most of the wavelet bases applied to the analysis of given data are not optimal. This means that the cascade does not hold in the microcanonical sense and so eq.~(\ref{eq:optimal}) cannot be used. In the following we will show that when the wavelet basis is relatively close to the optimal basis, the linear model proposed by Pottier et al., eq.~(\ref{eq:suboptimal}), is verified. Our proof is based on the QMF representation introduced in the previous subsection and it is focused on 1D signals for simplicity. The generalization of higher dimensions is straightforward.

First, let the optimal QMF be denoted by $(\vec{g},\vec{h})$. At the discretization level $j=0$, the signal corresponds to the vector $\vec{\beta}_0^{\mathrm{opt}}=(\ldots,s_{-1},s_0,s_1,\ldots)$. Let us consider now the Child and the Parent scale levels as the two next coarser dyadic levels, namely $j_\mathrm{C}=1$, $r_\mathrm{C} = 2$ pixels and $j_\mathrm{P}=2$, $r_\mathrm{P} = 4$ pixels (notice that the wavelet coefficients at levels $j \leq 0$ are all zero as discrete signals cannot vary inside their pixels, i.e., at levels finer than the discretization scale). This way, eq.~(\ref{eq:finest_wav_coeff}) is notated as:
\be
{\vec{\alpha}}_{C}^{\mathrm{opt}} \; = \; \mathbb{G} \cdot {\vec{\beta}}_0^{\mathrm{opt}}
\ee

\noindent
The approximation to the next level is given by:
\be
{\vec{\beta}}_1^{\mathrm{opt}} \; = \; \mathbb{H} \cdot {\vec{\beta}}_0^{\mathrm{opt}}
\ee

\noindent
from which the details at the coarser resolution (parent coefficients) can be deduced:
\bea
{\vec{\alpha}}_{P}^{\mathrm{opt}} &=& \mathbb{G} \cdot \vec{\beta}_1^{\mathrm{opt}}
\nonumber \\
                                  &=& \mathbb{G} \cdot \mathbb{H} \cdot \vec{\beta}_0^{\mathrm{opt}}
\eea

Owing to the fact that the QMF is optimal, at each location $k$ we can find an independent cascade variable $\eta_{k}$ such that:
\be
\alpha_{C, \, k}^{\mathrm{opt}} \; = \; \eta_k \, \alpha_{P, \left\lfloor {k}/{2} \right\rfloor}^{\mathrm{opt}}
\ee

\noindent
If we define now the matrix $\mathbb{N}$ formed by these cascade variables disposed on the diagonal, namely:

\be
\mathbb{N}_{k k'} \; = \; \eta_k \, \delta_{\left\lfloor {k}/{2} \right\rfloor k'}
\ee

\noindent
we have that the cascade relation between children and parent coefficients can be written for the child and parent detail vectors as follows:
\be
{\vec{\alpha}}_\mathrm{C}^{\mathrm{opt}} \; = \; \mathbb{N} \cdot {\vec{\alpha}}_\mathrm{P}^{\mathrm{opt}}
\label{eq:vector_optimal}
\ee

Let us now introduce a small perturbation on the optimal QMF; we will define a new, suboptimal QMF $(\vec{g}',\vec{h}')=(\vec{g}+\delta \vec{g},\vec{h}+\delta \vec{h})$ for small $\delta \vec{g}$ and $\delta \vec{h}$. The new child detail vector will be given by:
\bea
{\vec{\alpha}}_\mathrm{C} &=& (\mathbb{G}+\delta\mathbb{G}) \cdot \vec{\beta}_0
\nonumber \\
                          &=& {\vec{\alpha}}_\mathrm{C}^{\mathrm{opt}} + \delta\mathbb{G} \cdot \vec{\beta}_0
\nonumber \\
                          &=& \mathbb{N} \cdot {\vec{\alpha}}_\mathrm{P}^{\mathrm{opt}} + \delta\mathbb{G} \cdot \vec{\beta}_0
\label{eq:child_suboptimal}
\eea

\noindent
Notice that we have made the assumption $\vec{\beta}_0=\vec{\beta}_0^{\mathrm{opt}}$ as both are identified with the signal itself at its discretization scale.
The next coarser approximation vector is:
\bea
\vec{\beta}_1 &=& (\mathbb{H} + \delta \mathbb{H}) \cdot \vec{\beta}_0
\nonumber \\
              &=& \vec{\beta}_1^{\mathrm{opt}} + \delta \mathbb{H} \cdot \vec{\beta}_0
\eea

\noindent
Finally, the details at the next coarser resolution up to the first perturbation order are given by the following vector:
\bea
{\vec{\alpha}}_{P} &=& ({\mathbb{G} + \delta \mathbb{G}) \cdot \vec{\beta}_1^{\mathrm{opt}}} + \mathbb{G}\cdot \delta \mathbb{H} \cdot \vec{\beta}_0
\nonumber \\
                   &=& {\vec{\alpha}}_\mathrm{P}^{\mathrm{opt}} + (\delta \mathbb{G} \cdot \mathbb{H} + \mathbb{G} \cdot \delta \mathbb{H}) \cdot \vec{\beta}_0
\label{eq:parent_suboptimal}
\eea

\noindent
Combining eq.~(\ref{eq:child_suboptimal}) and eq.~(\ref{eq:parent_suboptimal}) we obtain:
\be
\vec{\alpha}_{C} \; = \; \mathbb{N} \cdot {\vec{\alpha}}_\mathrm{P} \: + \: \left[ \delta \mathbb{G} - \mathbb{N}\cdot (\delta \mathbb{G} \cdot \mathbb{H} + \mathbb{G} \cdot \delta\mathbb{H}) \right] \cdot \vec{\beta}_0
\label{eq:pre_vector_linear}
\ee

\noindent
Defining now ${\vec{\alpha}}_0$ as:
\be
{\vec{\alpha}}_0 \; \equiv \; \left[ \delta \mathbb{G} - \mathbb{N}\cdot (\delta \mathbb{G} \cdot \mathbb{H} + \mathbb{G} \cdot \delta\mathbb{H}) \right] \cdot \vec{\beta}_0
\label{eq:alpha0}
\ee

\noindent
when substituted in eq.~(\ref{eq:pre_vector_linear}) we obtain the vector version of the linear model, eq.~(\ref{eq:suboptimal}), introduced in \cite{Pottier.2008}, namely:
\be
{\vec{\alpha}}_{C} \; = \; \mathbb{N} \cdot {\vec{\alpha}}_\mathrm{P} \: + \: {\vec{\alpha}}_0
\label{eq:vector_linear}
\ee

\noindent
According to our derivation we can now make some remarks about the variables $\eta_0$ and $\alpha_0$ appearing in the linear model. First, the variable $\eta_0$ is an actual cascade variable, distributed according to the same statistics, and up to the first order it is independent from the parent coefficient in the suboptimal basis. Second, the variable $\alpha_0$ is much smaller than the term $\eta_0 \, \alphaP$ and is only relevant for small values of $\alphaP$. We cannot say much about the statistical distribution of $\alpha_0$, not even whether it is independent or not from the other term. However, it is reasonable to think that this variable is governed by the fluctuations due to the mixing of the different terms in the definition of $\alpha_0$ (see eq.~(\ref{eq:alpha0})) and the arbitrary character of the perturbations $\delta \mathbb{G}$ and $\delta \mathbb{H}$. This fact allows to consider this variable independent from $\eta_0 \, \alphaP$, as the experiences in \cite{Pottier.2008} confirm.

\subsection{Optimization strategies}

The results in the previous subsection show that the amplitude of $\alpha_0$ (the optimality degree) varies continuously under perturbations on the wavelet. Hence, an optimization strategy based on successive corrections of the wavelet would lead to the actual optimal wavelet, provided that the initial guess is not too far away from the optimality.

As seen in Section~\ref{sec:cascades} all cascade variables $\eta$ are equally distributed, independent of the wavelet basis from which they are derived, and their moments can be retrieved from $\tau_p$. In addition, the expectation value of $|\eta|$ is fixed due to translational invariance \cite{Turiel.2000a,Turiel.2000c}: $\langle |\eta| \rangle = 2^{-{d}/{2}}$ in an arbitrary dimension $d$; $\langle |\eta| \rangle = \frac{1}{\sqrt{2}}$ for 1D signals. According to the linear model, eq.~(\ref{eq:suboptimal}), the expectation value of $|\tilde{\eta}|$ is:
\be
\langle |\tilde{\eta}| \rangle = \langle | \eta_0 + \alpha_0 \, \alphaP^{-1} | \rangle
\ee

\noindent
Let us explore the two asymptotic limits. If the wavelet is optimal then $\alpha_0=0$ so:
\be
\langle |\tilde{\eta}| \rangle = \langle |{\eta}_0| \rangle  = \langle |{\eta}| \rangle
\ee

\noindent
In the opposite case, for a highly non-optimal wavelet we will have that $\alpha_0/\alphaP \gg \eta_0$ and taking $\alpha_0$ independent of $\alphaP$ we would obtain that:
\be
\langle |\tilde{\eta}| \rangle = \langle |\alpha_0| \rangle \langle |\alphaP|^{-1} \rangle = q \langle |{\eta}| \rangle
\ee

\noindent
where $q = \langle |\alphaP|\rangle \langle |\alphaP|^{-1} \rangle$, which by Jensen's inequality \cite{Rudin.1987} is greater than one: $q>1$, for any random variable $\alphaP$. For an intermediate case, the preceding two regimes are combined. If $p$ is the proportion of the range of values of $\alphaP$ for which $\eta_0 > \alpha_0/\alphaP$ and $(1-p)$ is its complementary, we roughly have that:
\be
\langle |\tilde{\eta}| \rangle \; \approx \; p \, \langle |\eta|\rangle \: + \: (1-p) \, q \, \langle |\eta | \rangle
\ee

\noindent
Hence, in any instance $\langle |\tilde{\eta}| \rangle \geq \langle |{\eta}| \rangle$ and $\langle |\tilde{\eta}| \rangle = \langle |{\eta}| \rangle$ for the optimal wavelet only. We normalize this quantity to define the optimality degree $Q$ as:
\be
Q = \frac{\langle |\tilde{\eta}| \rangle}{\langle |{\eta}| \rangle}
\label{eq:defQ}
\ee

\noindent
which is $Q \geq 1$, and $Q=1$ for the optimal wavelet only. $Q$ is a monotonic function of the amplitude of $\alpha_0$ (which in fact measures the deviation from the optimal case), so that $Q$  not only evidences the optimal wavelet case (when $Q=1$) but it actually ranks suboptimal wavelets by their respective deviation from optimality.

An alternative approach would consist in analyzing the degree of independence between $\tilde{\eta}$ and $\alphaP$. As stated in Section~\ref{subsec:microcascade}, independence between these variables is an indicator of the optimality of the wavelet. This can be expected, as having $Q>1$ implies correlation between $\tilde{\eta}$ and $\alphaP$, and correlation implies statistical dependence.
In this case decorrelation ($Q=1$) implies independence also, as $Q=1$ implies optimality and optimality implies independence. In fact, $\tilde{\eta}$ and $\alphaP$ are negatively correlated in suboptimal cases ($Q>1$), and uncorrelated only for the optimal wavelet:
\be
Q = \frac{\langle |\tilde{\eta}| \rangle}{\langle |{\eta}| \rangle} = 1 - \frac{\mathrm{Cov}(|\tilde{\eta}|, |\alphaP|)}{\langle |\alphaC| \rangle}
\ee

\noindent
A standard measure of statistical dependence is the mutual information. Therefore, the mutual information between $\tilde{\eta}$ and $\alphaP$, $I = I(\tilde{\eta}, \alphaP)$, could also measure the degree of optimality of a wavelet. However, the advantage of using $Q$ instead of $I$ comes from the fact that $Q$ is less numerically sensitive to sampling size than $I$. The main problem with the practical calculation of the mutual information is that it is very data demanding (see the estimation of uncertainties in \ref{app:infoconv} and the numerical study in the next section). Hence, when only small and short datasets are available, $Q$ is more convenient as indicator of the optimality degree.

\section{Results}
\label{sec:results}

Now we want to show in practice the theoretical results given in the previous section, namely the validity of the linear model, eq.~(\ref{eq:vector_linear}), and the performance of our measures of optimality, $Q$ and $I$.
We have generated synthetic signals according to a given cascade process and with a prefixed optimal wavelet basis.
The cascades are generated by first calculating the wavelet coefficients through eq.~(\ref{eq:optimal}) for dyadic scale steps, and then generating the signal from these wavelet coefficients, eq.~(\ref{eq:wv_expansion}), with the chosen wavelet basis.
The multiplicative variable $\eta$ is a random variable following a given cascade distribution without horizontal correlations, i.e., it follows Benzi et al.'s model \cite{Benzi.1993a}. As distribution for the cascade variable $\eta$ we have chosen the log-Poisson distribution, which has been proposed in many different physical systems \cite{Dubrulle.1994,She.1994,Turiel.2000a}. Hence, we have chosen a translationally invariant log-Poisson characterized by having a most singular manifold of dimension $D_\infty=0$ and singularity exponent $h_\infty=-\frac{1}{2}$, which is a realistic choice of parameters \cite{Turiel.1998,Turiel.2000a}.
See the \ref{app:connection} for a description of the log-Poisson distribution and parameters.

\begin{figure}[htb]
\begin{center}
\leavevmode
\includegraphics[width=4cm]{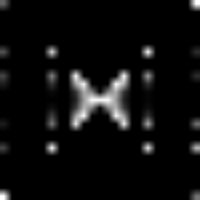}
\includegraphics[width=4cm]{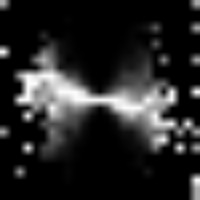}
\\ [0.05cm]
\includegraphics[width=4cm]{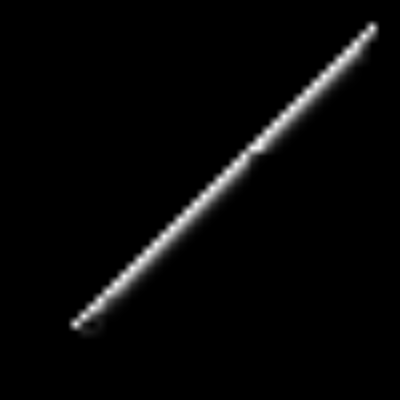}
\includegraphics[width=4cm]{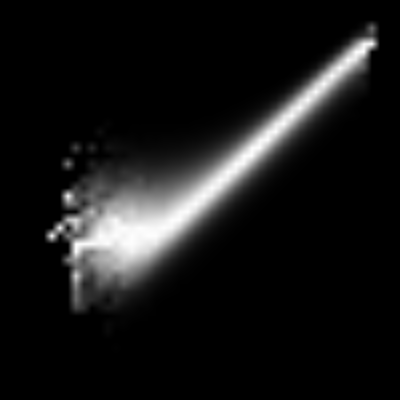}
\end{center}
\caption{Joint histograms of $\alphaC$ vs. $\alphaP$ (top) and $\ln |\alphaC|$ vs. $\ln |\alphaP|$ (bottom) for synthetic cascade data generated with the Coiflet-1 wavelet and analyzed with the Coiflet-1 wavelet (left figures, the optimal case) and Battle-Lemari\'e-6 wavelet (right figures, a non-optimal case). Each histogram column has been normalized so that it corresponds to the probability distribution function of the vertical-axis variable conditioned to a given horizontal-axis value. Values range $-0.125$ to $0.125$ in both axes (top figures) and $-32$ to $1$ in both axes (bottom figures).
The analyzed data are a single series of very high resolution (67 108 864 points) and the histograms are defined by a grid of $25\times25$ bins (top) and of $50\times50$ bins (bottom), smoothed with a cubic spline to enhance presentation.
The generating cascade process, eq.~(\ref{eq:dyadic_cascade}), is a log-Poisson of parameters $D_\infty = 0$ and $h_\infty = -\frac{1}{2}$ (see \ref{app:connection} for a detailed description of the process).}
\label{fig:illushisto}
\end{figure}

Regarding the linear model, it has been derived by perturbative analysis. In Figure~\ref{fig:illushisto} we validate this model in practice, for a very long series of 67 108 864 points. Figure~\ref{fig:illushisto} top shows the probability density function of the child coefficient $\alphaC$ conditioned by a given value of the parent coefficient $\alphaP$. In Figure~\ref{fig:illushisto} top left the analysis wavelet is the optimal wavelet and in Figure~\ref{fig:illushisto} top right the analysis wavelet is a suboptimal wavelet. First, we can observe that for any value of the parent coefficient, the child coefficient is symmetrically distributed $\rho(\alphaC | \alphaP) = \rho(-\alphaC | \alphaP)$, what means that $\langle \eta_0 \rangle = \langle \alpha_0 \rangle = 0$; this also implies $\rho(\alphaC | \alphaP) = \rho(\alphaC | -\alphaP)$. We also observe that the standard deviation of the child coefficient conditioned by a value of the parent coefficient depends hyperbolically on it, as predicted by the linear model, namely:
\be
\sigma_{\alphaC|\alphaP} = \sqrt{\langle \alphaC^2 | \alphaP \rangle} = \sqrt{A \, \alphaP^2 + B}
\ee

\noindent
where the constants $A$ and $B$ are given by the linear model: $\langle \eta_0^2 \rangle = A$ and $\langle \alpha_0^2 \rangle = B$. For the optimal wavelet, $A = \langle \eta^2 \rangle$ and $B=0$, so that $\eta_0$ coincides with $\eta$.
Additional evidence is furnished by the conditioned histograms of logarithms of the parent and child coefficients, i.e., the conditional probability of $\ln |\alphaC|$ for a given value of $\ln |\alphaP|$, which is shown in Figure~\ref{fig:illushisto} bottom. As before, the bottom left histogram corresponds to the optimal case while the bottom right histogram is a suboptimal case.
The absolute values fold the top  histograms to the first quadrant while the logarithms balance the kurtotic distributions of the wavelet coefficients. When the series is analyzed with its optimal wavelet, the histogram exhibits a perfectly straight maximum-probability line and small dispersion around this line. In contrast, when the series is analyzed with a suboptimal wavelet the histogram bends on the left to a horizontal line.
This bending is in agreement with the linear model, eq.~(\ref{eq:vector_linear}), as the term $\alpha_0$ becomes dominant when $\alphaP$ is too small. The two asymptotic limits can be easily obtained from eq.~(\ref{eq:vector_linear}):
when the value of the parent coefficient $\alphaP$ is large, in $\ln |\alphaC| = \ln |\eta_0 \, \alphaP| + \ln \left| 1 + \frac{\alpha_0}{\eta_0 \, \alphaP} \right|$ the second term becomes irrelevant,
so that $\ln |\alphaC| \approx \ln |\alphaP| + \ln |\eta_0|$. When the value of the parent coefficient $\alphaP$ is small, in $\ln |\alphaC| = \ln |\alpha_0| + \ln \left| 1 + \frac{\eta_0 \, \alphaP}{\alpha_0} \right|$ the second term rapidly becomes irrelevant, so that $\ln |\alphaC| \approx \ln |\alpha_0|$. Not only the asymptotes, but also the central behavior is the one given by the model, as the line of maximum-probability of the histogram fits a shape:
\be
\ln |\alphaC|_{\mathrm{m.p.}} = \ln \left( |\alpha_0|_{\mathrm{m.p.}} + |\eta_0|_{\mathrm{m.p.}} \, \exp \ln |\alphaP| \right)
\ee

\noindent
where m.p. stands for \emph{maximum probable}, i.e., these values are the probability maxima of their respective distributions. In addition, the amplitude of the fluctuations of $\alpha_0$ is larger than that of $\eta_0$. That is why the left side of the histogram shows large dispersion that is reduced as $\ln |\alphaP|$ grows and tends to that of the optimal case in the right side.

In a more extensive test, we have used 24 standard wavelets of very different families. These are: Haar, Daubechies (orders 2 to 10), Coiflet (orders 1 to 5), Symmlet (orders 4 to 8) and Battle-Lemari\'e (spline wavelets) (orders 1, 2, 3 and 6). Notice that Haar and Daubechies-1 coincide, while Symmlet 1 to 3 also coincide with Daubechies 1 to 3 respectively, and for that reason we have not repeated them (see \cite{Mallat.1999} for a description of these wavelet bases). For each wavelet, we have generated 64 series of 4096 points, which is a quite realistic size.
Hence, we have generated 24 ensembles of series and each wavelet is optimal in an ensemble. For a given ensemble, we have processed it with the same 24 wavelet bases. That is, for each ensemble we have tried its optimal basis and 23 non-optimal bases.
We have hence performed $24 \times 24 = 576$ different tests to check the validity of the theoretical results presented before.

In Figure~\ref{fig:cross_histo} we present the joint histograms of
$\ln |\alphaC|$ vs. $\ln |\alphaP|$, obtained from the different
ensembles when they are analyzed with the 24 bases, arranged in a
tabular form. By construction, the histograms on the diagonal of this
table correspond to the case in which the ensemble is analyzed with
its optimal wavelet, and hence these histograms exhibit the same
optimal behavior seen in Figure~\ref{fig:illushisto}, bottom left. In
contrast, when an ensemble is analyzed with a suboptimal wavelet the
histogram bends on the left to a horizontal line, as in
Figure~\ref{fig:illushisto}, bottom right. As the optimal and
analyzing wavelets become more different, the amplitude of the term
$\alpha_0$ increases and hence the extension of the horizontal line in
the joint histogram becomes longer.

\newpage

\begin{widetext}

\begin{figure}[htb]
\begin{center}
\includegraphics[width=15cm]{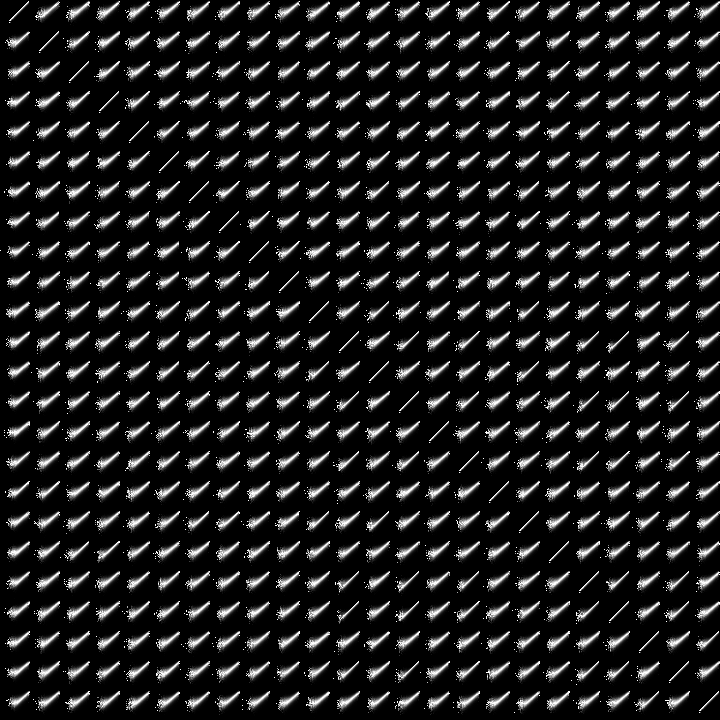}
\end{center}
\caption{Joint histograms of $\ln \alphaC$ vs. $\ln \alphaP$ for synthetically generated cascade process data. The histograms have been arranged as in Table~\ref{tab:results}, i.e., generation wavelet (rows) vs. analysis wavelet (columns), so that the main diagonal corresponds to the optimal wavelet cases. Each histogram column has been normalized so that it corresponds to the probability distribution function of $\ln \alphaC$ conditioned to a given value of $\ln \alphaP$ in the horizontal axis. Values range $-20$ to $1$ in both axes, the same for all the histograms. The analyzed ensembles correspond to 64 series of 4096 points each and the generating cascade process is a log-Poisson of parameters $D_\infty = 1$ and $h_\infty = -\frac{1}{2}$. Each histogram has $30\times30$ boxes.}
\label{fig:cross_histo}
\end{figure}

\begin{table}[htb]
{\tt \tiny
\begin{tabular}{|@{\,}c@{\,}|@{}c@{}c@{}c@{}c@{}c@{}c@{}c@{}c@{}c@{}c@{}c@{}c@{}c@{}c@{}c@{}c@{}c@{}c@{}c@{}c@{}c@{}c@{}c@{}c@{}|}
\hline
&Haar&Dau2&Dau3&Dau4&Dau5&Dau6&Dau7&Dau8&Dau9&DauA&Coi1&Coi2&Coi3&Coi4&Coi5&Sym4&Sym5&Sym6&Sym7&Sym8&BLS1&BLS2&BLS3&BLS6\\ \hline
\sidew{Haar\,\,}&\cola{1.00}&\colc{5.88}&\cold{6.69}&\cold{9.19}&\cold{8.40}&\cold{7.68}&\colc{5.63}&\colc{4.87}&\cold{7.42}&\cold{8.77}&\cold{7.12}&\colc{5.11}&\cold{7.30}&\cold{11.3}&\cold{8.15}&\cold{6.11}&\colc{4.86}&\cold{7.61}&\cold{8.18}&\cold{8.06}&\colc{5.44}&\cold{7.50}&\colc{5.94}&\cold{7.37}\\
&\cola{0.00}&\cold{0.13}&\colc{0.12}&\colc{0.12}&\cold{0.13}&\colc{0.12}&\colc{0.12}&\colc{0.12}&\cold{0.13}&\cold{0.14}&\cold{0.13}&\colc{0.12}&\colc{0.12}&\colc{0.12}&\colc{0.12}&\cold{0.13}&\colc{0.12}&\colc{0.12}&\cold{0.13}&\colc{0.12}&\colc{0.12}&\cold{0.13}&\colc{0.12}&\cold{0.13} \\ \hline
\sidew{Dau2\,\,}&\cold{8.81}&\cola{1.00}&\cold{22.9}&\cold{6.26}&\cold{8.92}&\cold{8.17}&\cold{11.8}&\cold{6.43}&\colc{5.13}&\colc{5.10}&\cold{9.91}&\cold{8.84}&\cold{9.66}&\cold{45.8}&\cold{12.2}&\colc{5.95}&\colc{4.28}&\cold{7.43}&\cold{7.22}&\cold{19.5}&\cold{8.42}&\cold{7.47}&\cold{6.38}&\cold{7.70}\\
&\colc{0.12}&\cola{0.00}&\colc{0.11}&\colc{0.12}&\cold{0.13}&\cold{0.13}&\colc{0.12}&\colc{0.12}&\colc{0.12}&\cold{0.13}&\cold{0.15}&\colc{0.12}&\cold{0.13}&\colc{0.12}&\colc{0.12}&\colc{0.12}&\colc{0.10}&\cold{0.13}&\cold{0.13}&\colc{0.12}&\colc{0.12}&\colc{0.12}&\colc{0.12}&\cold{0.13}\\ \hline
\sidew{Dau3\,\,}&\cold{6.53}&\cold{7.31}&\cola{1.00}&\colc{4.69}&\cold{9.96}&\cold{6.86}&\cold{14.6}&\cold{8.63}&\cold{7.04}&\colc{4.80}&\cold{14.3}&\cold{14.0}&\cold{7.42}&\cold{8.54}&\colc{5.28}&\cold{6.63}&\cold{8.41}&\cold{8.59}&\colc{3.83}&\cold{8.83}&\cold{8.24}&\cold{6.91}&\cold{7.01}&\cold{6.63}\\
&\colc{0.11}&\colc{0.12}&\cola{0.00}&\colc{0.11}&\cold{0.13}&\cold{0.14}&\cold{0.13}&\cold{0.13}&\colc{0.12}&\colc{0.12}&\colc{0.12}&\cold{0.13}&\cold{0.13}&\cold{0.13}&\cold{0.13}&\colc{0.12}&\cold{0.13}&\cold{0.13}&\colc{0.11}&\colc{0.12}&\cold{0.13}&\colc{0.12}&\cold{0.13}&\colc{0.12}\\ \hline
\sidew{Dau4\,\,}&\cold{9.10}&\cold{8.59}&\colc{4.97}&\cola{1.00}&\cold{21.8}&\colc{5.64}&\cold{11.4}&\cold{13.1}&\cold{9.09}&\colc{4.95}&\cold{10.6}&\cold{8.69}&\cold{6.70}&\colc{5.27}&\cold{6.25}&\cold{8.49}&\cold{38.6}&\cold{6.59}&\colc{3.35}&\colc{5.86}&\cold{8.68}&\cold{7.10}&\cold{7.63}&\cold{7.57}\\
&\colc{0.12}&\colc{0.12}&\colc{0.11}&\cola{0.00}&\colc{0.11}&\cold{0.13}&\cold{0.14}&\cold{0.13}&\cold{0.13}&\colc{0.12}&\colc{0.12}&\colc{0.12}&\cold{0.13}&\colc{0.12}&\colc{0.12}&\cold{0.13}&\colc{0.12}&\colc{0.12}&\colc{0.09}&\cold{0.13}&\colc{0.12}&\cold{0.13}&\cold{0.13}&\cold{0.13}\\ \hline
\sidew{Dau5\,\,}&\cold{9.04}&\cold{8.14}&\colc{5.88}&\colc{4.02}&\cola{1.00}&\colc{3.75}&\colc{5.86}&\cold{9.35}&\cold{12.2}&\cold{7.94}&\colc{6.00}&\colc{5.64}&\cold{6.95}&\cold{10.3}&\cold{7.73}&\cold{6.85}&\cold{8.78}&\cold{9.28}&\cold{6.85}&\cold{6.58}&\cold{6.31}&\cold{11.7}&\colc{5.24}&\cold{6.38}\\
&\cold{0.13}&\cold{0.13}&\cold{0.13}&\colc{0.11}&\cola{0.00}&\colc{0.11}&\cold{0.13}&\cold{0.14}&\cold{0.13}&\cold{0.13}&\cold{0.13}&\colc{0.12}&\cold{0.13}&\colc{0.12}&\cold{0.13}&\colc{0.12}&\cold{0.13}&\cold{0.13}&\colc{0.12}&\colc{0.11}&\colc{0.12}&\cold{0.13}&\colc{0.12}&\cold{0.13}\\ \hline
\sidew{Dau6\,\,}&\colc{5.57}&\cold{7.12}&\cold{6.58}&\cold{7.60}&\colc{3.65}&\cola{1.00}&\colc{4.02}&\cold{6.40}&\cold{8.41}&\cold{34.0}&\cold{6.05}&\cold{7.92}&\colc{4.41}&\colc{4.87}&\cold{10.6}&\colc{4.67}&\cold{6.06}&\cold{11.2}&\cold{6.51}&\cold{11.4}&\cold{6.79}&\cold{6.53}&\cold{6.35}&\colc{5.96}\\
&\cold{0.13}&\cold{0.14}&\cold{0.13}&\colc{0.12}&\colc{0.11}&\cola{0.00}&\colc{0.12}&\cold{0.13}&\cold{0.13}&\cold{0.13}&\colc{0.12}&\colc{0.12}&\colc{0.12}&\colc{0.12}&\cold{0.13}&\colc{0.12}&\cold{0.13}&\colc{0.12}&\cold{0.13}&\colc{0.12}&\colc{0.12}&\colc{0.12}&\colc{0.12}&\cold{0.13}\\ \hline
\sidew{Dau7\,\,}&\cold{10.2}&\cold{12.7}&\cold{12.8}&\cold{9.84}&\cold{7.72}&\colc{3.87}&\cola{1.00}&\cold{6.33}&\cold{8.00}&\cold{11.9}&\colc{4.63}&\colc{5.76}&\colc{4.62}&\cold{6.16}&\cold{6.73}&\cold{6.05}&\colc{5.77}&\colc{5.92}&\cold{7.55}&\cold{22.3}&\cold{6.71}&\cold{17.4}&\cold{6.12}&\cold{21.3}\\
&\cold{0.13}&\colc{0.12}&\cold{0.14}&\cold{0.13}&\cold{0.13}&\colc{0.11}&\cola{0.00}&\colc{0.11}&\cold{0.13}&\cold{0.13}&\cold{0.13}&\cold{0.13}&\colc{0.10}&\colc{0.12}&\cold{0.13}&\colc{0.12}&\colc{0.12}&\colc{0.11}&\cold{0.13}&\colc{0.12}&\colc{0.12}&\cold{0.13}&\colc{0.12}&\colc{0.12}\\ \hline
\sidew{Dau8\,\,}&\cold{6.64}&\colc{5.25}&\cold{6.19}&\cold{8.13}&\cold{10.8}&\cold{7.16}&\colc{5.71}&\cola{1.00}&\cold{9.82}&\cold{34.6}&\cold{7.38}&\cold{7.41}&\cold{7.18}&\cold{6.18}&\cold{11.4}&\cold{6.86}&\cold{6.24}&\cold{12.1}&\cold{7.85}&\colc{5.88}&\cold{7.81}&\cold{8.14}&\cold{7.79}&\cold{9.05}\\
&\cold{0.13}&\cold{0.13}&\cold{0.13}&\cold{0.14}&\cold{0.13}&\cold{0.13}&\colc{0.12}&\cola{0.00}&\colc{0.12}&\cold{0.13}&\colc{0.12}&\cold{0.13}&\colc{0.12}&\cold{0.13}&\cold{0.13}&\cold{0.13}&\colc{0.12}&\colc{0.11}&\colc{0.12}&\colc{0.12}&\cold{0.13}&\cold{0.14}&\cold{0.13}&\cold{0.13}\\ \hline
\sidew{Dau9\,\,}&\cold{21.3}&\cold{8.34}&\cold{8.11}&\cold{7.16}&\cold{6.56}&\cold{16.8}&\cold{8.50}&\colc{4.78}&\cola{1.00}&\cold{9.04}&\cold{8.17}&\cold{12.2}&\cold{8.51}&\cold{6.43}&\colc{5.23}&\cold{6.89}&\cold{8.32}&\cold{23.4}&\cold{7.06}&\cold{11.8}&\cold{11.7}&\colc{5.95}&\cold{7.76}&\cold{8.44}\\
&\cold{0.13}&\cold{0.13}&\colc{0.12}&\cold{0.13}&\cold{0.14}&\cold{0.13}&\cold{0.13}&\colc{0.12}&\cola{0.00}&\colc{0.12}&\cold{0.13}&\cold{0.14}&\cold{0.13}&\cold{0.14}&\cold{0.13}&\cold{0.13}&\colc{0.12}&\colc{0.12}&\colc{0.12}&\cold{0.13}&\cold{0.14}&\cold{0.13}&\cold{0.13}&\cold{0.14}\\ \hline
\sidew{DauA\,\,}&\cold{29.6}&\colc{4.95}&\cold{6.30}&\cold{8.67}&\cold{6.26}&\cold{15.9}&\cold{16.8}&\colc{5.85}&\colc{3.78}&\cola{1.00}&\cold{7.84}&\cold{12.8}&\cold{8.25}&\cold{7.09}&\colc{4.88}&\cold{9.75}&\cold{6.84}&\cold{7.82}&\colc{4.66}&\cold{7.99}&\cold{9.80}&\cold{9.54}&\cold{7.59}&\cold{7.09}\\
&\cold{0.13}&\cold{0.13}&\colc{0.12}&\colc{0.12}&\cold{0.13}&\cold{0.13}&\cold{0.13}&\cold{0.13}&\colc{0.11}&\cola{0.00}&\cold{0.14}&\cold{0.14}&\cold{0.13}&\cold{0.14}&\colc{0.12}&\cold{0.14}&\cold{0.13}&\cold{0.14}&\colc{0.12}&\cold{0.14}&\cold{0.14}&\cold{0.13}&\cold{0.14}&\cold{0.13}\\ \hline
\sidew{Coi1\,\,}&\cold{7.78}&\cold{10.1}&\cold{18.1}&\cold{6.30}&\cold{7.18}&\cold{9.07}&\colc{4.67}&\colc{5.30}&\cold{7.70}&\cold{9.89}&\cola{1.00}&\cold{11.4}&\colb{2.63}&\cold{6.98}&\cold{8.70}&\cold{6.59}&\cold{8.44}&\colc{3.02}&\cold{8.54}&\cold{6.14}&\cold{12.2}&\cold{8.40}&\colc{5.73}&\colc{5.51}\\
&\cold{0.13}&\cold{0.14}&\colc{0.12}&\colc{0.12}&\cold{0.13}&\colc{0.12}&\colc{0.12}&\cold{0.13}&\cold{0.13}&\cold{0.13}&\cola{0.00}&\cold{0.13}&\colc{0.09}&\cold{0.13}&\cold{0.13}&\cold{0.13}&\cold{0.13}&\colc{0.10}&\colc{0.12}&\cold{0.13}&\colc{0.12}&\colc{0.12}&\cold{0.13}&\colc{0.12}\\ \hline
\sidew{Coi2\,\,}&\cold{8.38}&\cold{6.25}&\cold{9.09}&\cold{9.42}&\cold{7.50}&\cold{6.29}&\colc{5.58}&\cold{6.38}&\cold{14.2}&\cold{15.3}&\cold{6.73}&\cola{1.00}&\cold{17.0}&\colb{2.05}&\cold{7.71}&\colb{2.62}&\cold{6.99}&\cold{7.29}&\cold{10.1}&\colb{2.23}&\cola{1.17}&\cold{44.1}&\cold{6.49}&\cold{6.26}\\
&\colc{0.12}&\colc{0.12}&\colc{0.12}&\colc{0.12}&\colc{0.11}&\colc{0.12}&\cold{0.13}&\cold{0.13}&\cold{0.13}&\cold{0.13}&\colc{0.12}&\cola{0.00}&\cold{0.13}&\colb{0.08}&\colc{0.12}&\colc{0.09}&\cold{0.13}&\cold{0.13}&\colc{0.12}&\colc{0.09}&\cola{0.04}&\colc{0.12}&\colb{0.08}&\cold{0.13}\\ \hline
\sidew{Coi3\,\,}&\cold{7.45}&\cold{10.0}&\cold{7.68}&\cold{10.6}&\colc{5.84}&\colc{4.56}&\cold{6.34}&\colc{5.12}&\cold{6.56}&\cold{7.38}&\colb{2.81}&\cold{6.69}&\cola{1.00}&\cold{6.75}&\cold{9.09}&\cold{7.09}&\cold{7.04}&\colb{2.35}&\cold{7.05}&\colc{5.88}&\cold{11.7}&\cold{6.36}&\cold{13.9}&\colc{4.87}\\
&\cold{0.13}&\colc{0.12}&\colc{0.12}&\colc{0.12}&\cold{0.13}&\colc{0.12}&\colc{0.09}&\colc{0.12}&\cold{0.13}&\cold{0.14}&\colc{0.10}&\cold{0.13}&\cola{0.00}&\cold{0.13}&\colc{0.12}&\cold{0.13}&\colc{0.12}&\colb{0.07}&\cold{0.13}&\cold{0.13}&\cold{0.13}&\colc{0.11}&\colc{0.12}&\colc{0.12}\\ \hline
\sidew{Coi4\,\,}&\cold{6.34}&\cold{7.85}&\cold{6.86}&\cold{7.58}&\colc{5.84}&\colc{5.76}&\colc{5.73}&\cold{6.36}&\cold{6.40}&\cold{7.65}&\colc{5.73}&\colb{2.06}&\cold{6.28}&\cola{1.00}&\cold{6.79}&\colc{3.22}&\colc{5.68}&\cold{16.8}&\cold{9.23}&\colb{1.86}&\colb{1.96}&\cold{9.38}&\colb{2.06}&\cold{7.18}\\
&\colc{0.12}&\colc{0.12}&\colc{0.12}&\colc{0.12}&\colc{0.11}&\colc{0.12}&\cold{0.13}&\cold{0.13}&\cold{0.13}&\cold{0.13}&\cold{0.13}&\colb{0.08}&\cold{0.13}&\cola{0.00}&\colc{0.12}&\colc{0.10}&\cold{0.13}&\cold{0.13}&\colc{0.12}&\colb{0.07}&\colb{0.06}&\cold{0.13}&\colb{0.06}&\cold{0.13}\\ \hline
\sidew{Coi5\,\,}&\cold{6.57}&\cold{8.36}&\cold{10.3}&\cold{6.42}&\cold{6.72}&\cold{7.27}&\cold{8.52}&\cold{6.09}&\cold{7.00}&\cold{6.72}&\cold{8.84}&\cold{9.85}&\cold{7.22}&\cold{11.9}&\cola{1.00}&\cold{15.8}&\cold{6.47}&\cold{8.84}&\colc{5.69}&\cold{6.80}&\cold{16.8}&\cold{20.7}&\cold{20.8}&\cold{9.33}\\
&\cold{0.13}&\colc{0.12}&\cold{0.13}&\cold{0.13}&\colc{0.12}&\cold{0.13}&\cold{0.13}&\cold{0.13}&\cold{0.13}&\cold{0.13}&\colc{0.12}&\colc{0.12}&\colc{0.12}&\colc{0.12}&\cola{0.00}&\colc{0.12}&\cold{0.13}&\cold{0.13}&\colc{0.12}&\colc{0.12}&\colc{0.12}&\cold{0.13}&\colc{0.12}&\cold{0.13}\\ \hline
\sidew{Sym4\,\,}&\cold{9.93}&\cold{11.9}&\cold{9.21}&\cold{6.49}&\cold{8.22}&\colc{4.73}&\colc{5.11}&\cold{7.21}&\cold{12.3}&\cold{7.29}&\colc{5.25}&\colc{4.15}&\cold{7.73}&\colb{2.84}&\cold{8.11}&\cola{1.00}&\cold{12.0}&\colc{5.55}&\cold{7.88}&\colc{3.46}&\colb{2.97}&\cold{6.26}&\cold{14.2}&\cold{6.58}\\
&\colc{0.12}&\colc{0.12}&\colc{0.12}&\colc{0.12}&\colc{0.12}&\colc{0.12}&\cold{0.13}&\cold{0.13}&\cold{0.13}&\cold{0.13}&\cold{0.13}&\colc{0.09}&\cold{0.13}&\colc{0.10}&\cold{0.13}&\cola{0.00}&\cold{0.13}&\cold{0.13}&\cold{0.13}&\colb{0.08}&\colc{0.09}&\colc{0.12}&\colc{0.10}&\cold{0.13}\\ \hline
\sidew{Sym5\,\,}&\cold{7.37}&\colc{4.95}&\cold{9.19}&\cold{14.8}&\cold{7.32}&\cold{6.19}&\cold{7.06}&\cold{6.19}&\colc{5.83}&\cold{6.14}&\cold{7.82}&\colc{5.82}&\cold{7.44}&\cold{6.27}&\cold{7.47}&\cold{9.90}&\cola{1.00}&\colc{6.00}&\cold{9.13}&\cold{6.94}&\cold{7.32}&\cold{8.81}&\cold{6.08}&\cold{11.1}\\
&\colc{0.12}&\colc{0.11}&\cold{0.13}&\cold{0.13}&\cold{0.13}&\colc{0.12}&\colc{0.12}&\colc{0.12}&\colc{0.12}&\cold{0.13}&\colc{0.12}&\colc{0.12}&\colc{0.11}&\cold{0.13}&\colc{0.12}&\colc{0.12}&\cola{0.00}&\colc{0.11}&\colc{0.12}&\colc{0.12}&\cold{0.13}&\cold{0.13}&\cold{0.13}&\cold{0.13}\\ \hline
\sidew{Sym6\,\,}&\cold{6.63}&\cold{7.73}&\cold{7.17}&\cold{10.7}&\cold{6.56}&\cold{7.94}&\colc{5.64}&\colc{4.26}&\cold{6.91}&\colc{5.90}&\colc{3.38}&\cold{8.96}&\colb{2.54}&\cold{39.2}&\cold{7.32}&\colc{5.54}&\colc{5.04}&\cola{1.00}&\cold{8.70}&\cold{6.01}&\cold{8.11}&\cold{6.21}&\cold{8.40}&\colc{5.38}\\
&\cold{0.13}&\cold{0.13}&\cold{0.13}&\colc{0.12}&\cold{0.13}&\cold{0.13}&\colc{0.11}&\colc{0.11}&\colc{0.12}&\cold{0.14}&\colc{0.09}&\cold{0.13}&\colb{0.08}&\cold{0.13}&\cold{0.13}&\cold{0.13}&\colc{0.11}&\cola{0.00}&\colc{0.12}&\cold{0.13}&\colc{0.12}&\colc{0.11}&\cold{0.13}&\colc{0.12}\\ \hline
\sidew{Sym7\,\,}&\cold{6.31}&\cold{6.04}&\colc{3.72}&\colb{2.90}&\cold{6.47}&\cold{6.91}&\cold{10.1}&\cold{6.46}&\colc{5.49}&\cold{8.23}&\cold{7.03}&\cold{11.5}&\cold{7.22}&\cold{21.3}&\cold{14.2}&\cold{6.82}&\cold{8.25}&\cold{8.60}&\cola{1.00}&\cold{8.43}&\cold{23.7}&\cold{6.24}&\cold{7.20}&\cold{7.05}\\
&\colc{0.12}&\cold{0.13}&\colc{0.10}&\colc{0.09}&\cold{0.13}&\cold{0.13}&\cold{0.13}&\colc{0.12}&\colc{0.12}&\colc{0.12}&\cold{0.13}&\cold{0.13}&\cold{0.13}&\colc{0.12}&\colc{0.12}&\colc{0.12}&\cold{0.13}&\colc{0.12}&\cola{0.00}&\cold{0.13}&\cold{0.13}&\colc{0.12}&\cold{0.13}&\colc{0.12}\\ \hline
\sidew{Sym8\,\,}&\cold{6.59}&\cold{7.59}&\cold{9.59}&\cold{7.13}&\cold{6.28}&\colc{4.72}&\colc{4.92}&\cold{18.5}&\cold{8.84}&\cold{6.77}&\cold{6.07}&\colb{2.53}&\colc{5.63}&\colc{3.62}&\cold{6.34}&\colb{2.92}&\cold{6.35}&\colc{4.82}&\cold{10.1}&\cola{1.00}&\colb{2.80}&\cold{8.96}&\colb{2.75}&\cold{6.62}\\
&\cold{0.13}&\colc{0.12}&\cold{0.13}&\colc{0.12}&\colc{0.12}&\colc{0.12}&\cold{0.13}&\cold{0.13}&\cold{0.13}&\cold{0.13}&\colc{0.12}&\colc{0.09}&\cold{0.13}&\colb{0.08}&\colc{0.12}&\colc{0.09}&\cold{0.13}&\cold{0.13}&\colc{0.12}&\cola{0.00}&\colc{0.09}&\colc{0.12}&\colc{0.09}&\cold{0.13}\\ \hline
\sidew{BLS1\,\,}&\cold{6.57}&\cold{6.78}&\cold{20.6}&\cold{6.97}&\cold{21.6}&\colc{4.78}&\cold{7.48}&\cold{6.87}&\cold{8.41}&\cold{6.35}&\colc{5.27}&\colb{2.43}&\cold{9.54}&\colb{1.95}&\cold{7.07}&\colc{4.16}&\cold{12.7}&\cold{8.60}&\cold{7.32}&\colb{2.44}&\cola{1.00}&\cold{6.59}&\colb{2.77}&\colc{5.18}\\
&\colc{0.12}&\colc{0.12}&\colc{0.12}&\colc{0.12}&\colc{0.11}&\colc{0.12}&\cold{0.13}&\cold{0.13}&\colc{0.12}&\cold{0.13}&\colc{0.12}&\cola{0.03}&\cold{0.13}&\colb{0.08}&\colc{0.12}&\colc{0.09}&\cold{0.13}&\cold{0.13}&\cold{0.13}&\colc{0.09}&\cola{0.00}&\cold{0.13}&\colc{0.09}&\cold{0.13}\\ \hline
\sidew{BLS2\,\,}&\cold{10.3}&\cold{11.8}&\colc{5.41}&\cold{9.95}&\cold{7.07}&\cold{6.45}&\cold{12.8}&\cold{8.66}&\cold{7.67}&\cold{7.45}&\cold{13.4}&\cold{14.2}&\colc{5.69}&\cold{6.70}&\colc{5.96}&\cold{6.94}&\cold{8.14}&\cold{9.08}&\cold{14.6}&\cold{6.92}&\cold{8.05}&\cola{1.00}&\cold{8.01}&\colb{2.68}\\
&\colc{0.12}&\colc{0.12}&\cold{0.13}&\colc{0.12}&\cold{0.13}&\cold{0.13}&\colc{0.12}&\cold{0.13}&\cold{0.13}&\colc{0.12}&\colc{0.12}&\cold{0.13}&\colc{0.12}&\cold{0.13}&\cold{0.13}&\cold{0.13}&\colc{0.12}&\colc{0.12}&\colc{0.12}&\cold{0.13}&\cold{0.13}&\cola{0.00}&\cold{0.13}&\colc{0.10}\\ \hline
\sidew{BLS3\,\,}&\cold{8.23}&\cold{6.90}&\cold{7.10}&\cold{7.66}&\cold{6.51}&\colc{4.91}&\cold{24.1}&\cold{7.02}&\cold{6.35}&\cold{8.48}&\cold{34.5}&\colc{3.58}&\cold{9.00}&\colb{2.18}&\cold{8.06}&\colc{3.47}&\colc{5.66}&\cold{6.72}&\cold{15.9}&\colc{3.36}&\colb{2.45}&\cold{6.38}&\cola{1.00}&\cold{12.0}\\
&\cold{0.13}&\colc{0.12}&\colc{0.12}&\colc{0.12}&\colc{0.12}&\colc{0.12}&\colc{0.12}&\cold{0.13}&\cold{0.13}&\cold{0.13}&\colc{0.12}&\colc{0.09}&\cold{0.13}&\colb{0.08}&\colc{0.12}&\colc{0.10}&\cold{0.13}&\cold{0.13}&\colc{0.12}&\colb{0.08}&\colc{0.10}&\cold{0.13}&\cola{0.00}&\colc{0.12}\\ \hline
\sidew{BLS6\,\,}&\cold{8.40}&\cold{8.15}&\cold{6.41}&\colc{5.87}&\cold{6.56}&\cold{8.52}&\cold{6.47}&\cold{18.8}&\cold{10.5}&\colc{5.29}&\cold{55.7}&\cold{6.16}&\colc{5.51}&\cold{8.46}&\colc{5.38}&\cold{7.17}&\cold{8.46}&\cold{7.57}&\cold{18.6}&\cold{7.22}&\cold{10.4}&\colb{2.72}&\colc{5.97}&\cola{1.00}\\
&\colc{0.12}&\colc{0.12}&\cold{0.13}&\cold{0.13}&\cold{0.13}&\cold{0.13}&\colc{0.12}&\cold{0.13}&\cold{0.13}&\cold{0.13}&\cold{0.13}&\cold{0.13}&\colc{0.12}&\cold{0.13}&\cold{0.13}&\cold{0.13}&\cold{0.13}&\colc{0.12}&\cold{0.13}&\cold{0.13}&\cold{0.13}&\colc{0.11}&\cold{0.13}&\cola{0.00}\\ \hline
\end{tabular}
}
\\
\begin{center}
{\tt \scriptsize
\begin{tabular}
{|@{\,}c@{\,}|@{}c@{}c@{}c@{}c@{}|}
\hline
$\ Q \ $ & \cola{  1.00 - 1.50  } & \colb{  1.50 - 3.00  } & \colc{  3.00 - 6.00  } & \cold{  > 6.00  } \\
\hline
$\ I \ $ & \cola{  0.00 - 0.04  } & \colb{  0.04 - 0.08  } & \colc{  0.08 - 0.12  } & \cold{  > 0.12  } \\
\hline
\end{tabular}
}
\end{center}
\caption{Summary of the $Q$ (upper side of the cell) and $I$ (lower side of the cell) optimality measures for synthetic cascade data. Each row corresponds to an ensemble generated with the wavelet written sideways at left (generation wavelet), while each column corresponds to the results obtained while analyzing these ensembles with the wavelet written at top (analysis wavelet). The ensembles correspond to 64 series of 4096 points each and the generating cascade process is a log-Poisson of parameters $D_\infty = 0$ and $h_\infty = -\frac{1}{2}$. Mutual information ($I$) is expressed in bits. Uncertainties of two sigmas are 0.0025 for $Q$ and 0.03 bits for $I$.}
\label{tab:results}
\end{table}

\end{widetext}

\clearpage

In Table~\ref{tab:results} we present the results of the mutual
information $I$ between $\tilde{\eta}$ and $\alpha_p$, and the $Q$
parameter as defined in eq.~(\ref{eq:defQ}) for the different
combinations of ensemble and analysis wavelet. As shown in the table,
only when the processing wavelet coincides with the optimal wavelet
the values of $I$ and $Q$ drop to 0 and 1, respectively, while for
other, non-optimal wavelets these values are always higher. This
proves that $Q$ has the same performance as $I$ to assess the
optimality of a wavelet basis, but the $Q$ parameter is less
statistically demanding. 

The $Q$ parameter is obtained by means of the average of
$\tilde{\eta}$ and so, according to the Central Limit Theorem, it
converges to its theoretical value with a standard deviation that
depends on the number of samples $N$ as $N^{-\frac12}$,
$\sigma_{\langle |\tilde{\eta}| \rangle} = \sigma_{|\tilde{\eta}|} \,
N^{-\frac12}$ (recall that the average in the denominator of $Q$,
$\langle |{\eta}| \rangle$, is theoretically fixed to
$\frac{1}{\sqrt{2}}$ due to translational
invariance). $\sigma_{|\tilde{\eta}|}$ depends on the wavelet and can
be analytically calculated for the optimal case only, which in fact is
the most interesting case as we want to have the error bar that
discriminates optimal from non-optimal wavelets. For the distribution
used here, log-Poisson with $D_\infty = 1$ and $h_\infty =
-\frac{1}{2}$, it is $\sigma_{|\eta|} = \sqrt{ 2^{-\frac{3}{4}} -
  \frac12 } = 0.31$, and so the standard deviation of $Q$ goes as
$0.62 \, N^{-\frac{1}{2}}$. As shown in \ref{app:infoconv}, the
estimation of the mutual information $I$ has also a standard deviation
depending on $N^{-\frac12}$, but the proportionality constant is
$\sqrt{\omega}$, which in our log-Poisson distribution is 5.66
bits. In addition, we do not take into account other sampling
uncertainties stated in \ref{app:infoconv} that do not depend on
$N$. The absolute uncertainty for $I$ is 12 times that of $Q$,
although their typical values are more than an order of magnitude
smaller. For these reasons, we have analyzed relative large ensembles
(64 series of 4096 points each) to show that $Q$ performs equally well
as $I$ for large ensembles, but $Q$ has the potential to be useful for
smaller ensembles. 

The results presented so far imply that the linear model is correct to describe the scaling relations in synthesized signals. We also know that this model is also correct for many wavelet basis on natural images \cite{Buccigrossi.1999,Wainwrigth.2001,Huang.1999,Schwartz.2001} and for satellite-derived chlorophyll maps \cite{Pottier.2008}. As a new example, we have processed the sequence of dye dispersed in 2D turbulence introduced in Section~\ref{sec:persistence}. This sequence consists of 81 snapshots, each one recorded on a 512$\times$ 512 gray-level image. The images were acquired with a CCD camera each 0.5 seconds and were calibrated so that the intensity level of each pixel is proportional to the dye concentration over that area. Some pre-processing is required in order to perform wavelet analysis of these data. First, data have a reduced dynamic range and acquisition noise is relevant, so we have reduced the resolution of images by averaging intensity values by blocks of 4$\times$ 4 pixels. We hence obtain a sequence 128 $\times$ 128 images with increased accuracy in the value at each pixel and increased signal-to-noise ratio. Second, it should be noticed that some areas of the image do not contain dye, specially during the early times of the sequence, and over those areas the wavelet coefficients vanish or take very small values. Henceforth, once the wavelet decomposition is obtained we have discarded those wavelet coefficients lying in areas at which dye concentration is negligible. The two stages of pre-processing considerably reduce the amount of available statistics, what is partially compensated by the fact of having 81 snapshots, although time correlation between snapshots is high and so the effective sampling size is moderate.

We have applied the 2D separable surrogates of the 24 wavelet bases
used so far, and we have studied the optimality of each basis. 2D
separable QMFs are formed by three different wavelet bases (also
called ``orientations'', denoted by horizontal (h), vertical (v) and
diagonal (d) details, \cite{Mallat.1999}), each one defining its own
cascade pyramid, in analogy to what was presented in
\cite{Turiel.2003a}. Hence, we must study the pyramid defined by each
orientation separately, what leads to define the parameters $Q_h$,
$Q_v$ and $Q_d$, and analogously for the mutual informations. Results
are summarized in Table~\ref{tab:Tabeling}.

The results indicate that the behavior of each orientation are not necessarily related, although a wavelet, Daubechies 5, attains a good performance for all orientations: attending to $Q$ parameter, this wavelet is the one closest to optimal for horizontal and vertical orientations, and is just 4.7\% above the best one for diagonal orientation. In addition, the values of the three mutual informations are in the lowest range for this wavelet. Notice that the value of mutual informations are not very significant for any of the wavelets due to the limited sampling size.

\newpage

\begin{widetext}

\begin{table}[htb]
{\tt \tiny
\begin{tabular}{|@{\,}c@{\,}|@{}c@{}c@{}c@{}c@{}c@{}c@{}c@{}c@{}c@{}c@{}c@{}c@{}c@{}c@{}c@{}c@{}c@{}c@{}c@{}c@{}c@{}c@{}c@{}c@{}|}
\hline
&Haar&Dau2&Dau3&Dau4&Dau5&Dau6&Dau7&Dau8&Dau9&DauA&Coi1&Coi2&Coi3&Coi4&Coi5&Sym4&Sym5&Sym6&Sym7&Sym8&BLS1&BLS2&BLS3&BLS6\\ \hline
\,H\,&\colb{2.79}&\colb{2.70}&\colb{2.64}&\colb{2.22}&\colb{1.81}&\colb{2.16}&\colc{3.32}&\colc{3.18}&\colb{2.59}&\colb{2.05}&\colc{3.04}&\colc{4.22}&\colc{5.01}&\colc{5.66}&\cold{10.0}&\colc{4.44}&\colc{4.03}&\colc{4.99}&\colc{4.00}&\cold{7.07}&\colb{2.71}&\colb{2.03}&\colb{2.10}&\colb{2.48}\\
&\colb{0.07}&\colb{0.04}&\cola{0.03}&\colb{0.04}&\cola{0.03}&\cola{0.01}&\cola{0.02}&\colb{0.06}&\cola{0.03}&\cola{0.03}&\cola{0.01}&\colb{0.05}&\cola{0.03}&\colb{0.05}&\colb{0.04}&\colb{0.05}&\colb{0.05}&\colb{0.06}&\colb{0.05}&\cola{0.02}&\colb{0.04}&\colb{0.04}&\cola{0.03}&\cola{0.03}\\ \hline
\,V\,&\colb{2.77}&\colb{2.71}&\colb{2.60}&\colb{2.17}&\colb{1.73}&\colb{2.12}&\colc{3.28}&\colc{3.14}&\colb{2.55}&\colb{1.99}&\colc{3.03}&\colc{4.08}&\colc{4.91}&\colc{5.41}&\cold{9.60}&\colc{4.36}&\colc{3.92}&\colc{4.89}&\colc{3.90}&\cold{6.73}&\colb{2.74}&\colb{2.02}&\colb{2.03}&\colb{2.51}\\
&\colb{0.08}&\colb{0.04}&\cola{0.03}&\colb{0.04}&\cola{0.03}&\cola{0.01}&\cola{0.02}&\colb{0.06}&\cola{0.03}&\cola{0.03}&\cola{0.01}&\colb{0.05}&\cola{0.03}&\colb{0.05}&\cola{0.03}&\colb{0.04}&\colb{0.05}&\colb{0.05}&\colb{0.05}&\cola{0.02}&\colb{0.04}&\cola{0.03}&\cola{0.03}&\cola{0.03}\\ \hline
\,D\,&\colb{2.85}&\colb{2.98}&\colc{3.02}&\colb{2.90}&\colb{2.96}&\colb{2.99}&\colc{3.00}&\colc{3.02}&\colb{2.90}&\colb{2.83}&\colc{3.04}&\colc{3.10}&\colc{3.04}&\colc{3.03}&\colc{3.01}&\colc{3.09}&\colc{3.02}&\colc{3.06}&\colc{3.00}&\colc{3.06}&\colc{3.09}&\colb{2.93}&\colb{2.93}&\colb{2.82}\\
&\colb{0.06}&\cola{0.02}&\cola{0.02}&\cola{0.02}&\cola{0.02}&\cola{0.01}&\cola{0.01}&\cola{0.03}&\cola{0.01}&\cola{0.01}&\cola{0.00}&\cola{0.01}&\cola{0.00}&\cola{0.01}&\cola{0.00}&\cola{0.01}&\cola{0.01}&\cola{0.01}&\cola{0.01}&\cola{0.00}&\cola{0.02}&\cola{0.02}&\cola{0.02}&\cola{0.01}\\ \hline
\end{tabular}
}
\\
\begin{center}
{\tt \scriptsize
\begin{tabular}
{|@{\,}c@{\,}|@{}c@{}c@{}c@{}c@{}|}
\hline
$\ Q \ $ & \cola{  1.00 - 1.50  } & \colb{  1.50 - 3.00  } & \colc{  3.00 - 6.00  } & \cold{  > 6.00  } \\
\hline
$\ I \ $ & \cola{  0.00 - 0.04  } & \colb{  0.04 - 0.08  } & \colc{  0.08 - 0.12  } & \cold{  > 0.12  } \\
\hline
\end{tabular}
}
\end{center}
\caption{Summary of the $Q$ (upper side of the cell) and $I$ (lower side of the cell) optimality measures for the dye dispersed in 2D turbulence sequence. Each row correspond to a different wavelet orientation, while each column corresponds to the results obtained while analyzing the sequence with the wavelet written at top (analysis wavelet). Mutual information ($I$) is expressed in bits.}
\label{tab:Tabeling}
\end{table}
\end{widetext}

\section{Conclusions}
\label{sec:conclusions}

In this paper we have discussed on the properties of optimal wavelet bases for the representation of multiplicative cascades. With the aid of optimal wavelet bases, any given signal originated by a cascade can be explicitly represented in terms of that multiplicative cascade.
s.
When the wavelet basis used in the analysis is suboptimal, the local cascade variables are poorly described. We have shown that the multiplicative process is perturbed by the inclusion of an additive, noise-like term, with an amplitude depending on the deviation from optimality of the studied basis.

We have then proposed to quantify the degree of optimality of a given basis with a simple descriptor $Q$, defined as the ratio of the first order moment of estimated cascade variable by the first order moment of the actual cascade variable. As this quantity is obtained from first-order moments, it is not demanding in data, and as any deviation implies an increase in the first order moment of the estimated cascade variable, the optimal wavelet is an absolute minimum of this quantifier. Hence, $Q$ can be used in any minimization strategy to derive the optimal wavelet.

To exemplify the derivations on the behavior of $Q$, we have used 24 different standard wavelets to analyze synthetic cascades generated with these same 24 wavelets. Our experiences reveal that $Q$ is more accurate than mutual informations in order to determine optimality in reduced datasets. In addition, as an example on real data, we have analyzed a sequence of dye dispersed in 2D turbulence and found that, among the 24 wavelets tested, Daubechies 5 is the closest to optimality.


With the help of the theory settled in this paper, we can undertake a more ambitious program of research. A natural future research line consists on implementing a continuous optimization strategy based on the descriptor $Q$, so that we could derive optimal wavelets of given databases of real signals without restricting the search to given families of wavelet bases. For each system we could hence prove if there exists such an optimal wavelet basis and even if it does not exist, we will be able to derive the best one. This would improve our knowledge on the dynamics of the studied systems.
Therefore, with the aid of optimal wavelets we could tackle problems such as data compression, forecast or inference of missing data, among others.

\section*{Acknowledgments}

We warmly thank Patrick Tabeling and Marie Caroline Jullien for the sequence of dye dispersed in 2D turbulence. This work is a contribution to OCEANTECH (PIF 2006 Project) and FIS2006-13321-CO2-01. O. Pont is funded by a Ph.D. contract from Generalitat de Catalunya.


\appendix

\section{Connection of the microcanonical cascade with the multifractal singularity spectrum}
\label{app:connection}

A signal $s$ is said to be multifractal in the microcanonical sense \cite{Turiel.2008} if an intensive functional $\epsilon_r$ acting on this signal (see Section~\ref{sec:cascades}) can be characterized by local scaling relations of the type:
\be
\epsilon_r(\vec{x}) \; = \; \alpha(\vec{x}) \: r^{h(\vec{x})} \: + \: o \left( r^{h(\vec{x})} \right)
\label{eq:multifractal}
\ee

\noindent
where the symbol $o \left(r^{h(\vec{x})}\right)$ means a term that is negligible in comparison with $r^{h(\vec{x})}$. The function that comprises the local properties of changes in scale, $h(\vec{x})$, is called the {\it singularity exponent} of the signal at the point $\vec{x}$ \cite{Turiel.2000a,Turiel.2008}. A signal verifying eq.~(\ref{eq:multifractal}) is said ``multifractal'' (in the microcanonical sense) because each value $h$ of singularity exponent is associated to a singularity component $F_h\equiv\{\vec{x} : h(\vec{x})=h\}$ of fractal character, with Hausdorff dimension $D(h)$. The function $D(h)$ is known as the singularity spectrum of the signal \cite{Falconer.1990}.

An interesting feature of the singularity spectrum is that although it is a geometrical feature of the multifractal, it completely defines the statistical properties of the cascade process. In fact, Parisi and Frisch \cite{Parisi.1985} proved that the knowledge of $D(h)$ granted the knowledge of the distribution of the cascade variables $\eta$ through the knowledge of the multiscaling exponents $\tau_p$, as expressed by eq.~(\ref{eq:anomalous}). In that case, $\tau_p$ is related to the singularity spectrum of the multifractal through a Legendre transform:
\be
\tau_p = \inf_h \left\{ ph + d - D(h) \right\}
\label{eq:PF85}
\ee

\noindent
which is known as the Parisi-Frisch formula and is the cornerstone of the canonical multifractal formalism. An interesting corollary of eq.~(\ref{eq:PF85}) is that when $D(h)$ is convex the Legendre transform can be inverted and hence $D(h)$ can be expressed as the Legendre transform of the multiscaling exponents $\tau_p$, namely:

\be
D_L(h) = \inf_p \left\{ ph + d - \tau_p \right\}
\label{eq:LegendreDh}
\ee

\noindent
The function $D_L(h)$ is the so-called Legendre singularity spectrum, which is a convex function of $h$ because Legendre transforms are always convex. If $D(h)$ is convex, $D(h)=D_L(h)$; if $D(h)$ is not convex, $D_L(h)$ will be its convex hull.

There is a more direct approach to $D(h)$ that can be used when the cascade variables are accessible and eliminates the necessity of imposing convex spectra $D(h)$. This approach consists in calculating the limit as $\kappa \to 0$ of the distribution of cascade singularity exponents. The cascade singularity exponents are defined as follows:
\be
h_{\kappa} \; = \; \log_{\kappa} \eta_{\kappa} \; = \; \frac{\ln \eta_{\kappa}}{\ln \kappa}
\label{eq:lim_to_h}
\ee

\noindent
where $\eta_{\kappa}$ is the multiplicative cascade variable that relates $\epsilon_r$ with $\epsilon_L$, $\kappa = r/L$, as in eq.~(\ref{eq:doteta}). The cascade singularity exponents represent the singularity exponents in the same sense of eq.~(\ref{eq:multifractal}) when they are obtained at the resolution level \cite{Turiel.2008}, i.e., when the scale ratio $\kappa$ is the one that compares the largest (whole-domain wide) scale $L$ with the smallest (resolution-level) scale $r$, meaning that $r << L$ or equivalently $\kappa \to 0$. As the singularity components $F_{h_\kappa}$ are of fractal character, the distribution of singularity exponents at a given observation scale behaves as \cite{Falconer.1990}:
\be
\rho(h_{\kappa}) \sim \kappa^{d-D(h_{\kappa})}
\ee

\noindent
with, as stated, $\kappa \to 0$. A direct obtaining of the $D(h)$ is hence possible through:
\be
\lim_{\kappa \to 0} \frac{\ln \rho(h_{\kappa})}{\ln \kappa} = d - D(h)
\label{eq:lim_to_Dh}
\ee

\noindent
where:
\be
h \equiv h_0 = \lim_{\kappa \to 0} h_{\kappa}
\label{eq:h_in_lim}
\ee

\noindent
{\it Lemma: The singularity spectrum derived according eq.~(\ref{eq:lim_to_Dh}) coincides with the Legendre spectrum, eq.~(\ref{eq:LegendreDh}), when the singularity spectrum is convex}

\noindent
{\bf Proof:} First, we define a random variable $h_\kappa$ such that $\eta_\kappa = \kappa^{h_\kappa}$, i.e.,
\be
h_{\kappa} = \frac{\ln \eta_{\kappa}}{\ln \kappa}
\ee

\noindent
As the cascade variable $\eta_\kappa$ is derived from a multifractal signal, the limit in eq.~(\ref{eq:lim_to_Dh}) exists and it is $d-D(h)$ (the Hausdorff spectrum of the signal) \cite{Turiel.2008}. Therefore, the distribution of $h_\kappa$ has a leading order $\kappa^{d-D(h_\kappa)}$ as follows: 
\be
\rho(h_\kappa) = A_\kappa \kappa^{d-D(h_\kappa)} + o (\kappa^{d-D(h_\kappa)})
\ee

\noindent
for small values of $\kappa$. Recalling here eq.~(\ref{eq:etamoments}) we have:
\be
\tau_p = \lim_{\kappa \to 0} \frac{\ln \langle \eta_\kappa^p \rangle}{\ln \kappa}
\ee

\noindent
We then expand it to find that:
\bea
\tau_p &=& \lim_{\kappa \to 0} \frac{1}{\ln \kappa} \ln \left( \int \! \mathrm{d} h_\kappa \, \kappa^{h_\kappa p} \rho(h_\kappa) \right)
\brn
       &=& \lim_{\kappa \to 0} \frac{1}{\ln \kappa} \ln \left( \int \! \mathrm{d} h_\kappa \, \kappa^{h_\kappa p} A_\kappa \kappa^{d-D(h_\kappa)} \right)
\brn
       &=& \lim_{\kappa \to 0} \inf_{h_\kappa} \{ h_\kappa p + d-D(h_\kappa) \}
\brn
       &=& \inf_{h} \{ h p + d-D(h) \}
\label{eq:fhiscoDh}
\eea

\noindent
where we used the saddle-point approximation. Notice that eq.~(\ref{eq:fhiscoDh}) is analogous to eq.~(\ref{eq:PF85}). Recalling that the inverse of a Legendre transform on convex functions is another Legendre transform, if we obtain now the Legendre spectrum, eq.~(\ref{eq:LegendreDh}), and assuming that $D(h)$ is convex we conclude $D_L(h)=D(h)$, q.e.d.

We will show now two examples of the lemma above, for two commonly used multiplicative processes, namely log-normal and log-Poisson processes. A log-normal process has the following distribution:
\be
\rho (\ln \eta_{\kappa}) = \frac{1}{\sqrt{2 \pi \sigma_{\kappa}^2}} e^{-\frac{1}{2} \left( \frac{\ln \eta_{\kappa} - \mu_{\kappa}}{\sigma_{\kappa}} \right)^2}
\label{eq:log-normal}
\ee

\noindent
Hence, the $\tau_p$ as defined in eq.~(\ref{eq:anomalous}) are given by:
\be
\tau_p = \frac{\mu_\kappa}{\ln \kappa} \, p + \frac{\sigma_\kappa^2}{2 \ln \kappa} \, p^2
\ee

\noindent
Let $h_m = \mu_{\kappa} / \ln \kappa$ and $\sigma_h^2 = -2 \sigma_{\kappa}^2 / \ln \kappa$ (remember that $\kappa < 1$), so eq.~(\ref{eq:LegendreDh}) leads to the singularity spectrum $D(h)$:
\be
D(h) = d - \left( \frac{h-h_m}{\sigma_h} \right)^2
\label{eq:Dhlog-normal}
\ee

\noindent
Let us show now that eq.~(\ref{eq:lim_to_Dh}) leads to the same expression. Notice that eq.~(\ref{eq:lim_to_h}) means that $\rho(h_\kappa) = -\ln \kappa \, \rho(\ln \eta_\kappa)$. Then, we substitute $\mu_{\kappa} = h_m \, \ln \kappa$ and $\sigma_{\kappa}^2 = -\sigma_h^2 \, \frac{\ln \kappa}{2}$ in eq.~(\ref{eq:log-normal}) to obtain:
\be
\frac{\ln \rho (h_{\kappa})}{\ln \kappa} = \left( \frac{h-h_m}{\sigma_h} \right)^2 + \frac{\ln \sqrt{\frac{-\ln \kappa}{\pi \sigma_h^2}}}{\ln \kappa}
\ee

\noindent
and the second term vanishes as $\kappa \to 0$ leading to eq.~(\ref{eq:Dhlog-normal}). It follows that eq.~(\ref{eq:lim_to_Dh}) holds.

The log-Poisson case is a little bit more elaborated due to the discrete-to-continuous passage. A log-Poisson process is defined as $\eta_{\kappa}=\kappa^{h_\infty}\beta^n$ with $n$ being a Poisson variable of parameter $\lambda$. Then the distribution of $\ln \eta_{\kappa}$ is:
\be
\rho (\ln \eta_{\kappa}) = \sum_{n=0}^{\infty} e^{-\lambda} \frac{\lambda^n}{n!} \, \delta (\ln \eta_{\kappa} - h_\infty \ln \kappa - n \ln \beta)
\label{eq:log-Poisson}
\ee

\noindent
which is discrete, i.e., it only takes nonzero values for some values of $\ln \eta_{\kappa}$. The parameter $h_\infty$ is the singularity exponent of the Most Singular Component (MSC) \cite{Turiel.2000a,Turiel.2006}, while the parameter $\lambda$ is related to the dimension of the MSC: $\lambda = (d-D_\infty)(-\ln \kappa)$ (both parentheses are always positive). It is also required that $0 < \beta < 1$.
After some simple algebra, it is obtained that $\tau_p$ are given by:
\be
\tau_p = p h_\infty + (d-D_\infty)(1-\beta^p)
\ee

\noindent
and through eq.~(\ref{eq:LegendreDh}) the singularity spectrum is:
\be
D(h) = D_\infty + (d-D_\infty) \, \omega(h) \, (1-\ln \omega(h))
\label{eq:DhlogPoisson}
\ee

\noindent
with
\be
\omega(h) = -\frac{1}{\ln \beta} \frac{h-h_\infty}{d-D_{\infty}}
\label{eq:w_h_poisson}
\ee

\noindent
Let us now apply eq.~(\ref{eq:lim_to_Dh}). From equations (\ref{eq:lim_to_h})~and~(\ref{eq:log-Poisson}), the $h_\kappa$ deviates from the most singular exponent $h_{\infty}$ in an integer number $n$ of contributions $\log_\kappa \beta$, namely
\be
h_\kappa = h_\infty + \underbrace{n \frac{\ln \beta}{\ln \kappa}}_{\Delta h_\kappa}
\label{eq:hlogPoisson}
\ee

\noindent
which give rise to a continuum of $h$ in the limit $(-\ln \kappa) \to \infty$. Let us now define a convenient auxiliary variable, $\omega(h_{\kappa})$, as
\bea
\omega(h_{\kappa}) &=& \frac{n}{\lambda}
\nonumber \\
 &=& \frac{1}{d-D_{\infty}} \frac{n}{(-\ln \kappa)}
\nonumber \\
 &=& -\frac{1}{\ln \beta} \frac{h_{\kappa}-h_\infty}{d-D_{\infty}}
\eea

\noindent
Notice that $\omega(h_{\kappa})$ is positive and proportional to $\Delta h_\kappa$. We now recall eq.~(\ref{eq:log-Poisson}) to obtain:
\be
\frac{\ln \rho (h_{\kappa})}{\ln \kappa} = \frac{-\lambda + n \ln \lambda - \ln n!}{\ln \kappa}
\ee

\noindent
Hence, according to eq.~(\ref{eq:lim_to_Dh}), the singularity spectrum is:
\be
D(h) = d -(d-D_\infty) + \lim_{\kappa \to 0} \frac{n \ln \lambda - \ln n!}{-\ln \kappa}
\ee

\noindent
Where $h = h_{\kappa \to 0}$ as in eq.~(\ref{eq:h_in_lim}). For any $h_\kappa$ different from $h_\infty$, i.e., $\Delta h_\kappa \neq 0$, when $\kappa$ goes to 0, $n$ grows accordingly, because $n$ is proportional to $(-\ln \kappa)$. So the limit $\kappa \to 0$ is equivalent to $n \to \infty$:
\be
D(h) = D_\infty + \lim_{n \to \infty} \frac{n \ln \lambda - n \ln n + n - \ln (\sqrt{2 \pi n}) }{-\ln \kappa}
\ee

\noindent
where we have used the Stirling approximation to expand $n!$. Recalling $(-\ln \kappa) = n \, ((d-D_{\infty}) \, \omega(h_\kappa) )^{-1}$ we have:
\be
D(h) \; = \; D_\infty + (d-D_{\infty}) \lim_{n \to \infty} (\ln \lambda - \ln n + 1) \, \omega(h_\kappa)
\ee

\noindent
which, as $\omega(h_\kappa) = n / \lambda$, leads to eq.~(\ref{eq:DhlogPoisson}).

\section{Convergence of the estimates of the mutual information}
\label{app:infoconv}

In this section we will calculate the standard deviation of the empirical estimates of the mutual information between two random variables. The mutual information between two variables $X$ and $Y$ is given by the following expression:
\be
I(X,Y) \; = \; H(X)+H(Y)-H(X,Y)
\ee

\noindent
where $H(X)$, $H(Y)$ stands for the marginal entropies and $H(X,Y)$ stands for the joint entropy. We will calculate the standard deviation on the estimate of $H(X,Y)$, the other two being completely analogous. The ideal joint entropy is given by:

\be
H(X,Y) \; = \; - \sum_{n,m} p_{nm} \ln p_{nm}
\ee

\noindent
We will express the entropy in nats instead of in bits for convenience in later calculations, so we use natural logarithms.

Let us now suppose that we sample the state space with a histogram of $B_x \times B_y$ boxes. We will assume that the sampling is efficient, so each box contains only one $p_{nm}$ at most. Void boxes, in case they exist, can directly be discarded as they do not change anything in the calculation, so without loss of generality we can assume that the sampling is perfect, and each box contains one and only one weight $p_{nm}$, so we can index boxes according to the weight index: $B_{nm}$. Notice however that $B_x$ and  $B_y$ are finite. We will first assume that they are large enough to make the contribution by the uncounted tails negligible.

For a sample of $N$ independent events we estimate $p_{nm}$ with the following statistical: 
\be
\hat{p}_{nm} \; = \; \frac{N_{nm}}{N}
\ee

\noindent
where $N_{nm}$ is the number of events happening to lie in box $B_{nm}$. The joint distribution of the variables $N_{nm}$ is a multinomial of order $N$ with $B_x \times B_y$ variables, each with probability $p_{nm}$. If $N$ is large enough we can disregard correlations and consider the distribution of each $N_{nm}$ as an independent binomial; notice however that this independence cannot be assumed when the weight estimates are summed up, as $\sum_{n,m} \hat{p}_{nm}=1$. Notice also that correlations would increase our estimate of uncertainty, so the value we are going to obtain should be considered a lower bound. Let us introduce a convenient representation for $\hat{p}_{nm}$:

\be
\hat{p}_{nm} \; = \; p_{nm} \: + \: \delta p_{nm}
\ee

\noindent
where we can express the deviation $\delta p_{nm}$ in the following way:
\be
\delta p_{nm} \; = \; \sqrt{\frac{p_{nm}(1-p_{nm})}{N}} \;\; \epsilon_{nm} \; \approx \; \sqrt{\frac{p_{nm}}{N}} \;\; \epsilon_{nm}
\ee

\noindent
where the random variable $\epsilon_{nm}$ is standardized (i.e., it has zero mean and unit variance).

The estimated joint entropy $\hat{H}(X,Y)$ is hence given by:
\be
\hat{H}(X,Y) \; = \; - \sum_{n,m} \hat{p}_{nm} \ln \hat{p}_{nm}
\ee

\noindent
Expanding this expression up to the first order in $\delta p_{nm}$ we have:
\bea
\hat{H}(X,Y) &\approx& H(X,Y) \: - \: \sum_{nm} p_{nm} \ln \left( 1 +
\frac{\delta p_{nm}}{p_{nm}} \right) 
\nonumber \\ 
& & -  \sum_{n,m} \delta p_{nm} \ln p_{nm} 
\\
             &\approx& H(X,Y) \: - \: N^{-\frac{1}{2}}
 \sum_{n,m}\epsilon_{nm} \: p_{nm}^{\frac{1}{2}} \: \ln p_{nm} \nonumber
\eea

\noindent
where we have used that $\sum_{n,m} \delta p_{nm}= \sum_{n,m} \hat{p}_{nm} - \sum_{n,m} p_{nm}=1-1=0$. We conclude that the deviation between the estimate of the joint entropy and its actual value is given by the following expression:
\bea
\delta H(X,Y) &\equiv& \hat{H}(X,Y) - H(X,Y)
\nonumber \\
              &\approx& \: - \: N^{-\frac{1}{2}} \sum_{n,m}\epsilon_{nm} \: p_{nm}^{\frac{1}{2}} \: \ln p_{nm}
\eea

\noindent
$\hat{H}(X,Y)$ converges to $H(X,Y)$ when $N$ goes to infinity. Hence, in order to compute the standard deviation of $\hat{H} (X,Y)$ we just need to compute that of $\delta H(X,Y)$. Now taking the variables $\epsilon_{nm}$ as independent (a first order approximation), we have:
\be
\delta H(X,Y) \; = \; \sqrt{ \frac{\omega_{XY}}{N}} \: \epsilon
\ee

\noindent
where $\epsilon$ is a standardized variable and $\omega_{XY}$ is given by:
\be
\omega_{XY} \; = \; \langle (\ln p_{nm})^2 \rangle
\ee

Analogous expressions arise for $\delta H(X)$ and $\delta H(Y)$, having their corresponding $\omega_{X}$ and $\omega_{Y}$ respectively. Hence, the deviation of the mutual information estimate $\delta I(X, Y)$ is given by the squared sum of the deviations of the joint and marginal entropies, with a global $\omega$ that is the sum of the joint $\omega_{XY}$ and marginal $\omega_X$ and $\omega_Y$. Thus we can estimate the minimum number of samples $N_0$ to attain a given accuracy level $\delta I(X, Y)$ according to the following expression:
\be
N_0 \; = \; \frac{\omega}{\delta I(X, Y)^2}
\ee

The dependence on the square of the accuracy level makes entropy estimation very demanding in data. For instance, to attain an accuracy of 0.1 bits ($\approx 0.07$ nats) we have $N_{0.1}\approx 200 \, \omega$; to attain an accuracy of 0.01 bits we need a sample 100 times larger, $N_{0.01} \approx 2\cdot 10^4 \, \omega$. For the case studied in section~\ref{sec:results}, 4096-point series generated with log-Poisson distribution of parameters $D_\infty = 0$ and $h_\infty = -\frac{1}{2}$, the computed value of $\omega$ is around $\omega = 15.4 \ \mathrm{nats}^2$.

As a final remark, notice that we have made important assumptions to derive this formula. The two most significant ones depend on the properties of the sampling using $B_x \times B_y$ boxes. First, we have assumed that we have properly sampled the histogram; second, we considered that the non-sampled tails do not significantly contribute to uncertainty. Concerning the first, we are assuming that the sample of the state space with $B_x \times B_y$ boxes is such that the associated weights $\{ p_{nm} \}$ give an accurate idea of the mutual information; for instance, if $X$ and $Y$ are independent then that with a good approximation $p_{nm}=p_n^x \, p_m^y$. Concerning the second, we need to assume that the excluded tails decay fast enough not to significantly alter the value of the entropies. These two contributions will increase the dispersion $\delta H$ estimated here in a way that does not depend on $N$, so the mutual information will never be decreased below a certain level even if $N$ goes to infinity. These sampling effects are absolutely depending on the distribution we are considering and hence no a priori bound can be given here.

\bibliography{paper}

\end{document}